\documentclass[12pt,letterpaper]{article}
\usepackage[a4paper, total={7in, 10in}, margin=1in]{geometry}

\usepackage{graphicx}
\usepackage{helvet}
\usepackage{authblk}
\usepackage{hyperref}
\usepackage{amsmath} 
\usepackage{amssymb} 
\usepackage{orcidlink} 
\usepackage{float}
\usepackage[utf8]{inputenc}
\usepackage{algorithm}
\usepackage{amssymb}
\usepackage{tabularx}
\usepackage{booktabs}
\usepackage{array}
\usepackage{geometry}
\usepackage{url}   
\usepackage{hyperref} 
\usepackage{adjustbox}
\usepackage{setspace}
\usepackage[noend]{algpseudocode} 
\usepackage[super,comma,sort&compress]  
   {natbib}
\usepackage{svg}
\usepackage{caption}
\usepackage{makecell}

\makeatletter
\renewcommand{\maketitle}{\bgroup\setlength{\parindent}{0pt}
\begin{flushleft}
  \textbf{\@title}
  
  \@author
\end{flushleft}\egroup}
\makeatother

\title{Alpha-Z divergence unveils further distinct phenotypic traits of human brain connectivity fingerprint}
\date{}

\author[1]{Md Kaosar Uddin }
\author[2]{Nghi Nguyen}
\author[1]{Huajun Huang}
\author[3]{Duy Duong-Tran}
\author[1,*]{Jingyi Zheng}

\affil[1]{Department of Mathematics and Statistics,Auburn University, Auburn, Alabama, USA}
\affil[2]{Faculty of Science, Vrije Universiteit Amsterdam, the Netherlands}
\affil[3]{Department of Biostatistics, Epidemiology and Informatics, University of
Pennsylvania, Philadelphia, Pennsylvania,USA}

\affil[*]{Correspondence: jzz0121@auburn.edu}

\begin{document}

\maketitle

\section*{SUMMARY}

The accurate identification of individuals from functional connectomes (FCs) is critical for advancing individualized assessments in neuro/psychiatric research. Traditional metrics such as Pearson correlation and Euclidean distance fail to capture the non-Euclidean geometry of FCs, while geodesic distances (e.g., affine-invariant, log-Euclidean) require task- and scale-specific regularization and degrade under high-dimensional conditions. To address these challenges, we propose a novel distance measure, the Alpha-Z Bures-Wasserstein divergence, a geometry-aware metric for functional connectome comparison.  Unlike prior methods, our approach does not require meticulous parameter tuning and maintains strong identification performance across multiple task conditions, scan lengths, and spatial resolutions. We compare our method with classical (e.g., Euclidean, Pearson) and state-of-the-art manifold-based distances (e.g., affine-invariant, log-Euclidean, Bures–Wasserstein) and assess how changing regularization strengths affects geodesic distance performance across Human Connectome Project tasks and parcellation scales. Our results demonstrate that the proposed method significantly improves the identification rates over traditional and existing geodesic distance measures, particularly when optimized regularization is applied, and notably in high-dimensional settings where matrix rank deficiencies degrade the performance of existing metrics. Furthermore, we validate the generalizability of our approach across different functional connectivity conditions, including resting-state and task-based fMRI, using multiple parcellation schemes. These findings establish Alpha-Z as a reliable and generalizable framework for functional connectivity analysis, enhancing sensitivity to cognitive and behavioral patterns and offering strong potential for individualized clinical neuroscience.

\section*{KEYWORDS}


Brain connectomics; FC fingerprinting; Alpha-Z divergence; Individual fingerprint;Full rank condition

\section*{INTRODUCTION}

Brain activity is commonly inferred indirectly by measuring fluctuations in the blood oxygenation level dependent (BOLD) signal through magnetic resonance imaging (MRI), a technique that tracks oxygen consumption in the brain \cite{bandettini1992bold, frahm1992bold, kwong1992bold, ogawa1990bold, ogawa1992bold}. Functional magnetic resonance imaging (fMRI) has emerged as the gold standard for capturing this activity, offering noninvasive insights into brain function. Functional connectivity (FC) between two brain regions is typically defined as the statistical relationship between their respective BOLD signals, most often measured using Pearson’s correlation coefficient \cite{bravais1846, galton1886}. These relationships are captured in symmetric correlation matrices, known as functional connectomes (FCs), which represent the structure of connectivity of the whole brain \cite{fornito2016,sporns2018}. FCs have become crucial tools in neuroscience, providing insights into how the brain’s network organization changes due to factors such as aging \cite{zuo2017}, cognitive abilities \cite{shen2017}, and neuropsychiatric disorders \cite{xu2022consistency,garai2024effect,xu2024topology,fornito2015,vandenheuvel2019}. In addition to these large-scale patterns, FCs have been shown to reveal consistent, individual-specific connectivity patterns, referred to as ``brain fingerprint'' \cite{abbas2020geff,abbas2023tangent,amico2019towards,amico2021toward,duong2021morphospace,chiem2022improving,duong2024homological,duong2024principled,duong2024preserving,duong2024principled_m,xu2022consistency,garai2023mining,song2024causality,asee_peer_48152}. These unique patterns are stable across repeat fMRI sessions (separated by days to weeks)  and across  different scanning conditions \cite{finn2015functional,amico2018,venkatesh2020comparing} , allowing for accurate identification of individuals from a large group . The reproducibility of brain fingerprints has made them invaluable in predicting behavior, cognitive function, and even susceptibility to mental health conditions, further highlighting their potential in clinical neuroscience and personalized medicine \cite{finn2015functional,amico2018,venkatesh2020comparing,xu2024topology}.

The study of FCs has significantly evolved, largely driven by advancements in neuroimaging and computational methods. Traditionally, Pearson’s correlation coefficient has been the primary method for comparing FCs \cite{finn2015functional}. This method, while straightforward, has several limitations, particularly its assumption of linearity and its inability to capture the non-Euclidean geometry inherent in FC data. Such limitations have been highlighted in several key studies\cite{finn2015functional} emphasizing the method’s limited accuracy in reliably distinguishing individual-specific connectivity patterns. Recognizing such shortcomings motivates the development of geometry-aware methods that better respect the underlying manifold structure of FC data.

\begin{figure}[H]
    \centering
    \includegraphics[width=\textwidth]{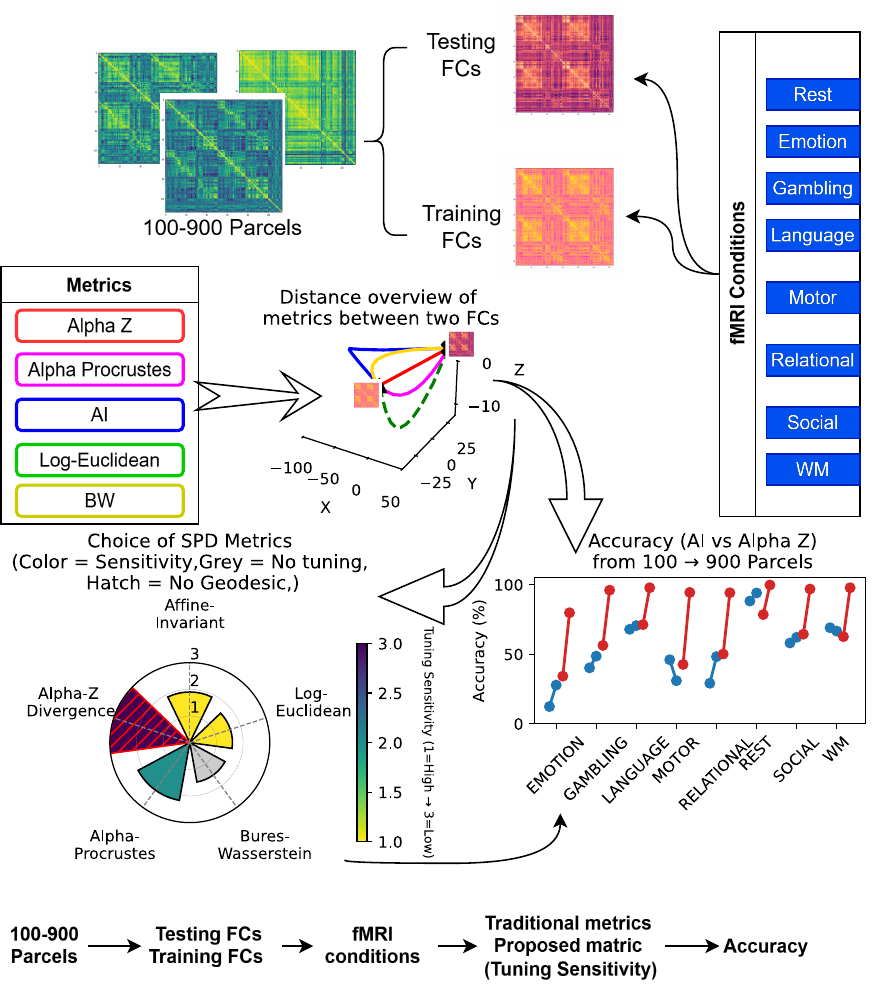}
    \caption{\textbf{A geometric overview of the metric domain explored in this study.} Training and testing FCs are generated from nine spatial granularities (100 – 900 parcels) and measured in eight fMRI tasks (blue column). The geometric behavior of metrics( the proposed Alpha-Z and Alpha-Procrustes, together with the widely used Affine-Invariant (AI), Log-Euclidean, and Bures–Wasserstein (BW)) between two FCs is showed in the central 3-D inset. The poler plot encodes overall  performance, bar color encodes how delicately the metric must be tuned (high = yellow, low = purple). Hatched fill marks metrics that lack a geodesic formulation, and gray marks indicates methods that require no hyper-parameter tuning. As we increase spatial granularity (100 → 900 parcels), the line chart  shows how much farther Alpha-Z can move along the manifold before distances collapse relative to each classical direction. The diagram highlights how both metric choice and hyper-parameter along with finer granularities demands drive identification performance, motivating to focus on Alpha-Z divergence as a promising direction for the functional connectomics analyses.
    }
    \label{fig:FCoverview}
\end{figure}

Recognizing the need for more sophisticated methods, a recent study \cite{venkatesh2020comparing} introduced the use of geodesic distance as a more accurate way to compare FCs. This method leverages the non-Euclidean geometry of the positive semidefinite cone, where FCs naturally reside. The introduction of geodesic distance represented a significant advancement, as it allowed for more precise measurement of differences between FCs by considering their curved manifold structure rather than treating them as flat, Euclidean objects. Geodesic distance was shown to significantly improve the identification rates of individual fingerprints, particularly when FCs were appropriately regularized to ensure they were positive definite and invertible. However, this approach, which involved  a fixed regularization parameter (e.g.,$\tau =1$), did not fully address the variability inherent across different datasets, brain parcellation methods, and scanning lengths. This limitation was further demonstrated to be highly dependent on specific dataset characteristics \cite{abbas2021geodesic}. A one-size-fits-all approach to regularization, therefore, would potentially diminish the accuracy of geodesic distance in capturing individual differences.

The superiority of geodesic distance over conventional metrics like Pearson-based correlations has been attributed to regularization techniques to ensure the invertibility of FC matrices \cite{venkatesh2020comparing,abbas2021geodesic}. However, this success was primarily produced under ``low-resolution'' scenarios, defined here as brain parcellation scales where the number of brain regions is smaller than the number of available time points, thus preventing rank deficiency. However, our current research specifically investigates ``high-resolution'' scenarios, characterized by parcellations where the number of brain regions significantly exceeds the number of available time points, inherently resulting in rank-deficient FC matrices. Under these high-resolution conditions, the performance of geodesic distance measures notably declines even with high tuning parameter. Notably, in a previous study \cite{abbas2021geodesic}, while a range of regularization values were investigated (including up to $\tau =10$) the identification rate generally declined as $\tau $ increased beyond certain optimal points. In this study, a smaller $\tau$ value was ultimately selected, \textit{i.e.}, $\tau =0.1$, to avoid the distortions that larger regularization values could introduce.
 
In general, Riemannian geodesic distances such as the affine‐invariant and log‐Euclidean have demonstrated clear advantages over flat, Euclidean measures by respecting the curved geometry of FCs and substantially boosting subject‐identification rates \cite{venkatesh2020comparing, abbas2021geodesic}. However, these approaches introduce new challenges: their reliance on a regularization parameter (\textit{e.g.}, \(\tau\)) requires careful, tasks and scale‐specific tuning; identification accuracy can drop sharply when the number of parcels exceeds the number of time points (\textit{i.e.}, in high‐resolution parcellations); and exhaustive searches for optimal regularization parameter values impose significant computational overhead. This outcome suggests that although selecting a small $\tau$ value helps preserve the original positions of the FC matrices within the manifold, the geodesic distance metric may still be suboptimal under certain conditions. These limitations motivate the exploration of alternative approaches that are both geometrically principled and robust to variations in resolution and task.


To address these issues, we introduce the Alpha-Z Bures–Wasserstein divergence \cite{dinh2021alpha}, a two-parameter extension of the standard BW distance that balances robustness and sensitivity across parcellation resolutions and task conditions. While \((\alpha)\) and ($z$) could in principle be tuned for each scenario, we demonstrate that a single fixed pair \((\alpha^*,z^*)\) performs well across all resolutions and tasks, eliminating the need for separate regularization searches and preserving high identification accuracy. We propose a novel approach that integrates advanced metrics, namely BW \cite{zheng2023barycenter}, Alpha Procrustes distance\cite{minh2022alpha} and Alpha-Z divergence\cite{dinh2021alpha}---into the comparison of FCs. We hypothesize that these new methods provide more accurate and robust individual fingerprints, particularly when combined with adaptive regularization techniques tailored to the specific characteristics of each dataset. We present evidence supporting this hypothesis through extensive comparisons across multiple cognitive tasks and parcellation granularity, demonstrating that our approach significantly enhances the precision of FC comparisons. We also systematically evaluated a range of distance metrics for FCs, considering both traditional (e.g., Euclidean, Pearson) and state-of-the-art manifold-based approaches (e.g., Affine-Invariant, Log-Euclidean, and BW, Alpha Procrustes) to support our findings.

The paper is structured as follows: Star methods describes in details about the datasets and preprocessing protocols employed in our analysis. It also introduces our methodology along with algorithm, including the mathematical foundations of the BW, alpha Procrustes distance and Alpha-Z divergence. In the result section, we present our experimental results, comparing the performance of the proposed methods against existing metrics. Finally, the discussion section explains the implications of our findings for future research and clinical applications in personalized medicine. Our conclusion demonstrates that by incorporating these advanced metrics, we can achieve a significant improvement in the precision of FC comparisons, paving the way for more individualized approaches to brain connectivity analysis.

\section*{RESULTS}
\subsection*{Comparative Connectome Identifiability Across Metrics, Tasks, and Spatial Resolutions}
The identification performance of various distance metrics in distinguishing individual FCs was systematically evaluated across eight fMRI tasks (Rest, Emotion, Gambling, Language, Motor, Relational, Social, and Working Memory (WM)) and a range of parcellation resolutions (100–900 regions). Identification rate\cite{finn2015functional}, defined as the percentage of correctly matched subjects based on their FCs, served as the primary performance metric. Here, we present the overall identification rates for manifold-based metrics(Log Euclidean, AI Distance, Alpha-Z Divergence, Alpha Procrustes Distance, and BW Distance), as well as  breaking these results down by cognitive tasks fMRI condition, highlighting key trends and relative metric strengths under each condition.

To further probe how spatial granularity and matrix conditioning interact with identifiability, we analyzed the impact of parcellation resolution on both identification rates and the rank (stability/invertibility) of the resulting symmetric positive definite (SPD) FC matrices. Figure \ref{fig:parcellation_rank} provides a comparative view of identification rates across spatial scale, revealing how finer parcellations can both enhance discriminability and challenge metric robustness as matrix rank increases.


\begin{figure}[H]
    \centering
    \includegraphics[width=\textwidth]{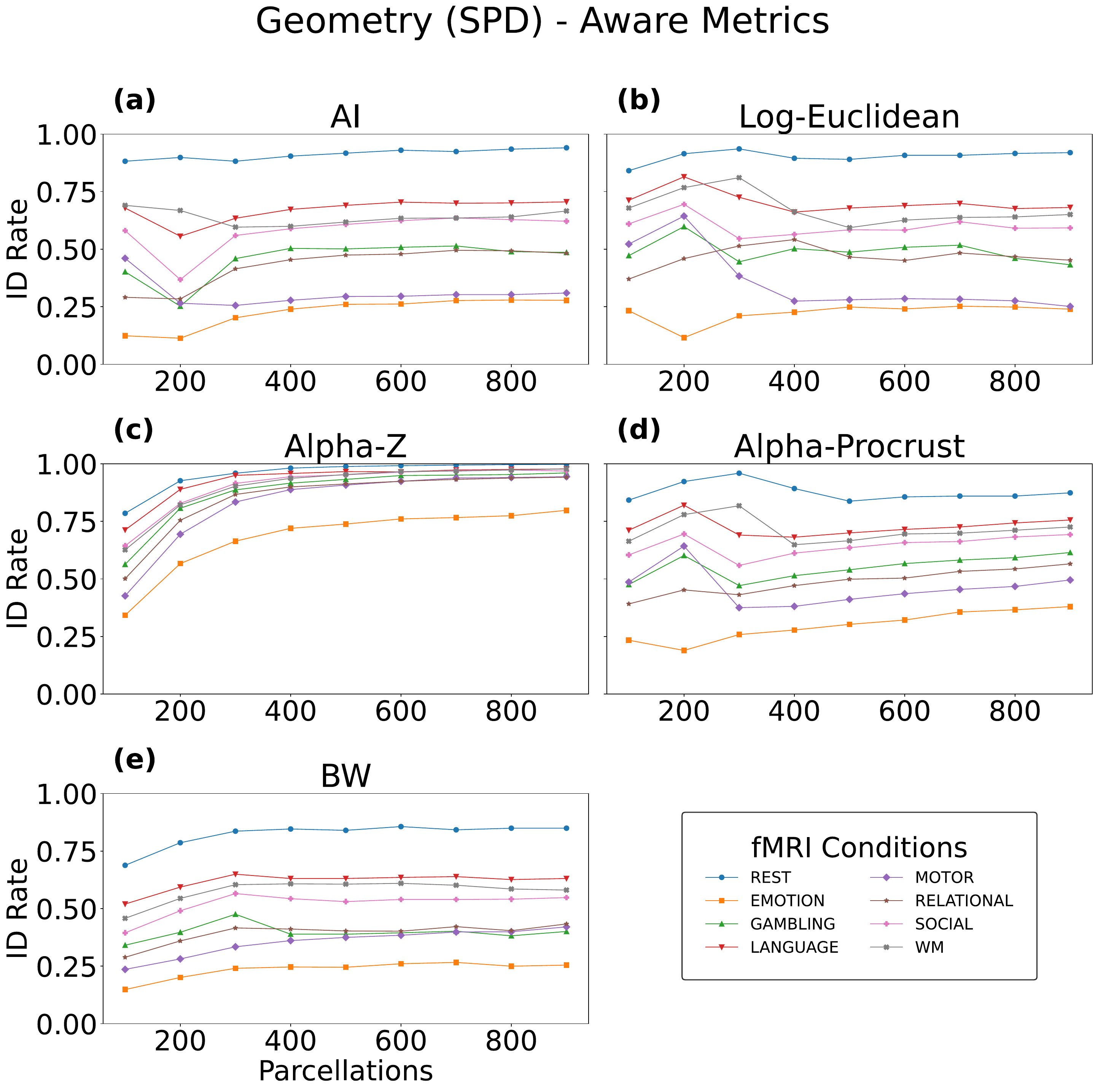}
    \caption{Impact of Different Distance Metrics on Identification Rates across parcellation resolutions and fMRI tasks. The five metrics evaluated are: Log Euclidean, AI Distance, Alpha-Z Divergence, Alpha Procrustes Distance, and BW Distance.}
    \label{fig:distancemetrics}
\end{figure}
\subsubsection*{Cross-Metric Identification Performance}
The performance of AI Distance (Fig.\ref{fig:distancemetrics}a) exhibits a more fluctuating pattern. Although it performs reasonably well at lower parcellation levels, the identification rate drops markedly beyond 300 regions for most tasks, including Emotion and Gambling, where performance does not exceed 50\%. At higher resolutions (600–900 parcellations), AI Distance shows signs of stabilization, particularly in tasks like Rest, but it still underperforms compared to other metrics, such as Alpha-Z Divergence. 

The Log Euclidean metric (Fig.\ref{fig:distancemetrics}b) demonstrates relatively high identification rates at lower parcellation resolutions (100–300 regions). However, a sharp decline is observed as the parcellation resolution increases, particularly for tasks like Motor and Gambling, where identification rates drop significantly beyond 400 parcellations. The accuracy rate curve follows the same trend as the other geodesic distance (AI). The performances remain below 60 \% for Motor, Gambling, and Relational for especially after 400 parcellations. This suggests that while Log Euclidean distance is effective for low-dimensional FC comparisons, it struggles to capture the complexity inherent in high-resolution parcellations. Even with regularization (as applied in AI distance), its performance does not recover.

Alpha-Z Divergence (Fig.\ref{fig:distancemetrics}c) consistently provides superior performance across all tasks and parcellation resolutions. Unlike the other metrics, Alpha-Z Divergence shows a steady increase in identification rates as the parcellations resolution increases, reaching high accuracy even at the maximum tested resolution (900 regions). All tasks except Emotion show particularly strong performance, maintaining high identification rates regardless of parcellation size. The robustness of Alpha-Z Divergence highlights its ability to handle the complex geometric structure of FCs, making it the most effective metric in high-dimensional settings.

The performance of Alpha Procrustes Distance (Fig.\ref{fig:distancemetrics}d) closely mirrors that of Alpha-Z Divergence, though with marginally lower identification rates across most tasks. This metric performs well across a wide range of parcellations, particularly for tasks such as Rest and Social, where it maintains high accuracy even at the highest resolutions. Its stability across varying resolutions underscores its suitability for high-dimensional FC comparisons. Therefore, it is a strong alternative to Alpha-Z Divergence.

BW Distance (Fig.\ref{fig:distancemetrics}e) performs moderately well, showing results comparable to AI Distance at lower parcellations but without the need for regularization. However, its identification rates for tasks like Emotion and Gambling remain relatively low, even as parcellation resolution increases. Despite this, Rest achieves reasonably high identification rates, suggesting that BW Distance may be a viable option for specific tasks where tuning parameter is not desirable. Nevertheless, its overall performance does not match that of Alpha-Z Divergence or Alpha Procrustes Distance in high-dimensional settings.


\subsubsection*{Task‐ and Performance‐Specific Trends}

\begin{figure}[H]
    \centering
    \includegraphics[width=\linewidth]{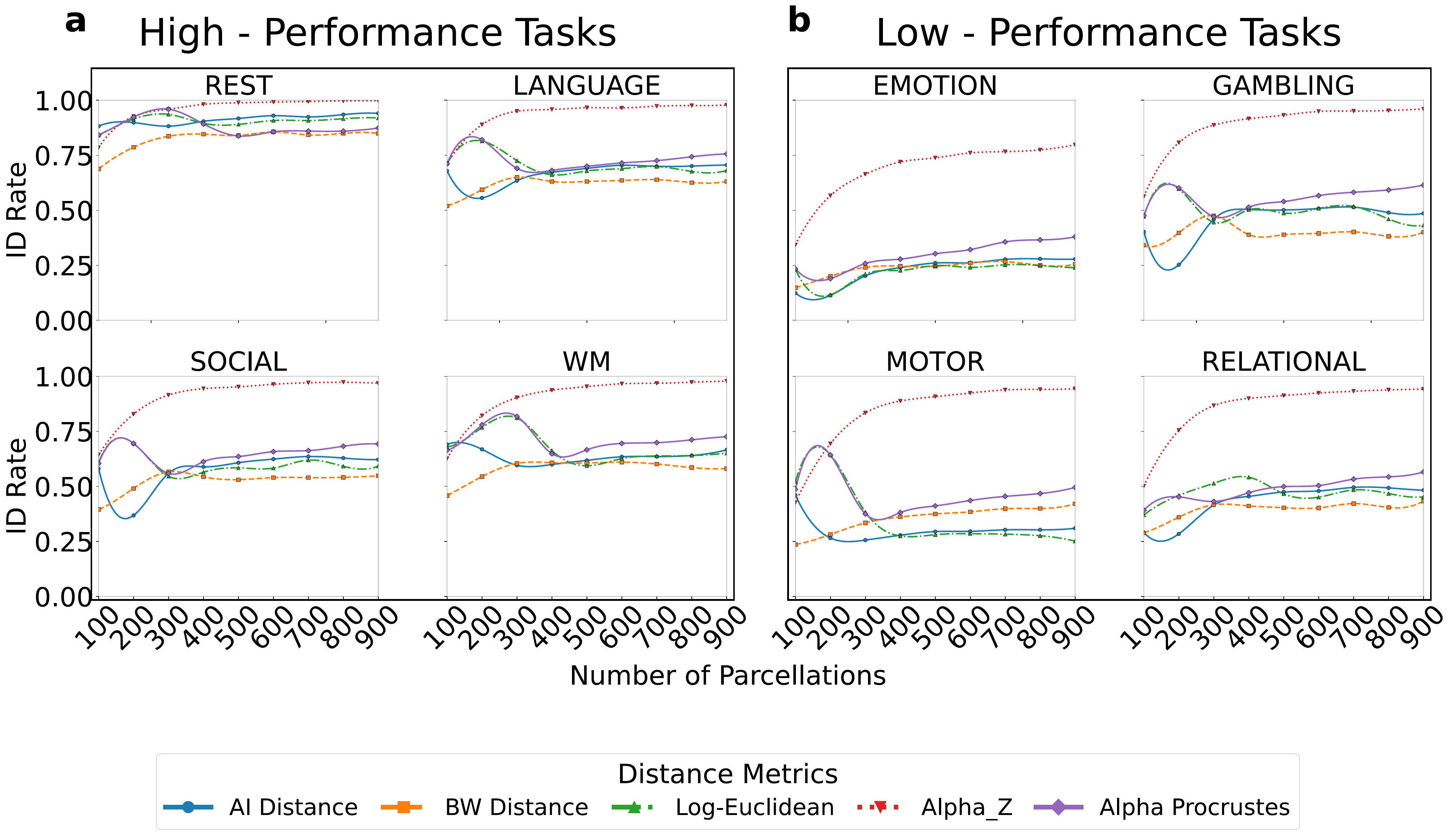}
    \caption{Identification rate (ID rate) across parcellation scales for four fMRI conditions with generally higher discriminative performance (Rest,Language, Social, and Working Memory (WM)) in panel \textbf{a} and  lower discriminative performance (Emotion, Gambling, Motor, Relational) in panel \textbf{b}, despite the default complex structure  of panel \textbf{b} tasks, the Alpha-Z distance consistently outperforms all other metrics, which shows Rest is the best performances tasks among all of the cognitive tasks with Alpha-Z divergence. 
    }
    \label{fig:TASKS}
\end{figure}

For the Rest task, as shown in the top-left panel of Figure \ref{fig:TASKS}a, Alpha-Z Divergence consistently outperforms the other metrics across all parcellation levels. Identification rates for this metric remain high even as the number of regions increases,approaching an accuracy rate of nearly 1.0 at all parcellation levels. Log Euclidean Distance and AI Distance show relatively stable but lower performance, while BW Distance and Alpha Procrustes Distance experience 2\% to 3\% decline in accuracy compared to AI and Log Euclidean Distance as resolution increases. This trend suggests that Alpha-Z Divergence more effectively captures the individual variation in the Rest task, maintaining high accuracy even as the data complexity increases.

In the Social task (bottom-left panel of Fig.\ref{fig:TASKS}a), similar trends are observed: Alpha-Z Divergence again outperforms all other metrics, maintaining identification rates above 0.9 across the entire range of parcellations. Unlike in the Rest task, Alpha Procrustes follows closely behind, also performing consistently well. AI Distance and Log Euclidean Distance provide moderate performance rates after 300 parcellations but still trail behind Alpha Procrustes, which highlights their limitations in capturing complex, high-dimensional FC structures. BW Distance performs worse than AI distance, although it does not require any tuning parameters.

For the Language task (top-right panel of Fig.\ref{fig:TASKS}a), the same pattern emerges. Alpha-Z Divergence achieves the highest accuracy across all resolutions, maintaining high accuracy even at 900 parcels. AI Distance and Log Euclidean Distance exhibit stable, despite lower, identification rates. BW Distance follows a similar pattern, with accuracy declining as dimensionality increases. Nevertheless, all distances maintain identification rates above 60\% across all parcellation levels.

In the Working Memory (WM) task (bottom-right panel of Fig.\ref{fig:TASKS}a), Alpha-Z Divergence and Alpha Procrustes again lead in performance. BW Distance, Log Euclidean and AI Distance perform reasonably well compared to the other tasks. AI Distance and Log-Euclidean exhibit a decrease and an increase in performance before reaching 300 parcels, respectively, but both show consistent performance at all higher resolutions. However, their inability to fully capture high-dimensional structure is once again evident.


As seen in the (top-left panel of Fig.\ref{fig:TASKS}b), the Emotion task reveals a clear separation between the metrics. Alpha-Z Divergence significantly outperforms the others, achieving approximately 80\% identification accuracy at all parcellation levels. In contrast, AI Distance, BW Distance, and Log Euclidean remain consistently low  performances, achieving around 30\% identification accuracy at all parcellation levels even when regularization is applied. Alpha Procrustes ranks second but still lags behind Alpha-Z Divergence. These results highlight the challenges faced by traditional metrics in capturing the complex, task-specific FC structures associated with emotional processing.

In the Gambling task (top-right panel of Fig.\ref{fig:TASKS}b), a similar trend is observed. Alpha-Z Divergence again outperforms the other metrics across all parcellations. Alpha Procrustes follows albeit a notable performance gap. BW Distance performs slightly worse than AI Distance and Log-Euclidean Distance. Both AI and Log-Euclidean distances follow a similar pattern for higher parcellation levels but exhibit significant variability at lower levels, failing to exceed 50 \% for all resolutions.

For the Motor task (bottom-left panel of Fig.\ref{fig:TASKS}b), Alpha-Z Divergence and Alpha Procrustes continue to dominate, maintaining high accuracy even at the highest parcellation levels. AI Distance and Log-Euclidean Distance yield moderate results, performing around 30 \%  for all parcellation levels. BW distance exhibits higher performance compared to AI Distance and Log Euclidean Distance, but could not come close to the Alpha Procrustes Distance. These results further reinforce the robustness of Alpha-Z Divergence in capturing FC patterns evoked by the motor tasks.

Finally, in the Relational task (bottom-right panel of Fig.\ref{fig:TASKS}b), Alpha-Z Divergence and Alpha Procrustes once again show the best performance across parcellations.Alpha-Z Divergence achieves high identification rates close to 0.9, which are the highest across all low-performance conditions. BW Distance proves minimally effective, with consistently low accuracy. AI Distance and Log-Euclidean deliver moderate identification rates, with Log-Euclidean slightly outperforming AI Distance at certain resolutions. However, the accuracy rates for AI Distance and Log Euclidean Distance remain below 50 \% across all parcellations.



\subsubsection*{Parcellation Granularity and Performance Trends}

\begin{figure}[H]
    \centering
    \includegraphics[width=\textwidth]{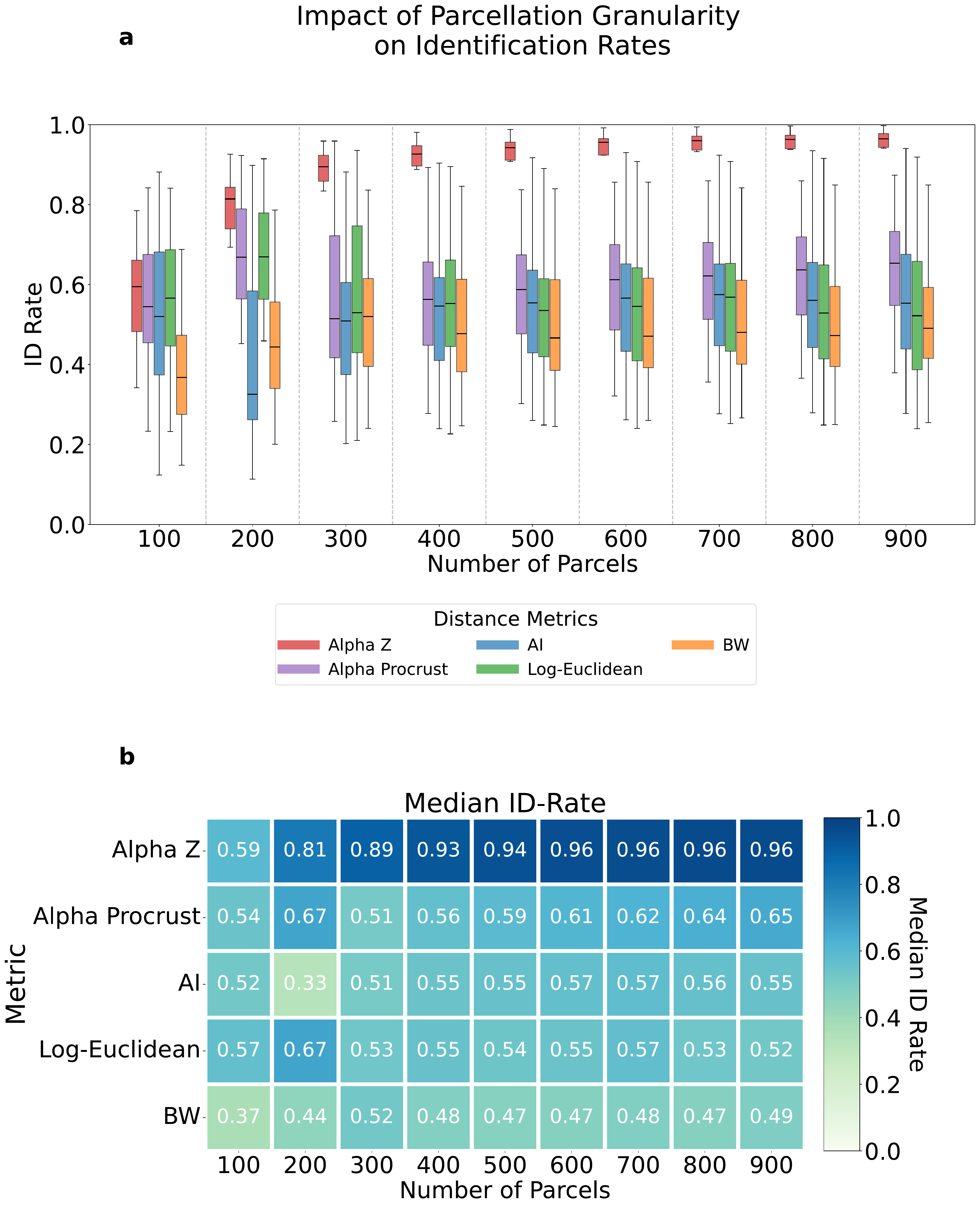}
    \caption{ Impact of parcellation granularity on identification rates across eight cognitive tasks (Rest, Emotion, Gambling, Language, Motor, Relational, Social, and Working Memory (WM)) using five methods: Alpha-Z, Alpha-Procrustes, Affine-Invariant (AI), Log-Euclidean, and Bures-Wasserstein (BW). Each portion that divided by dash line corresponds to a different parcellation scale (100 to 900 regions), illustrating how identification performance varies with spatial resolution in (panel \textbf{a}). And Median identification rate also showed in (panel \textbf{b}) across the parcellation resulations.The Alpha-Z method consistently achieves the highest identification rates across all parcellations levels, exploring its scalability and robustness to changes in network granularity.
    }
    \label{fig:parcellation_rank}
\end{figure}

As parcellation resolution increases from 100 to 900 regions, the box-and-whisker distributions in Fig. \ref{fig:parcellation_rank}(a) show a clear metric-dependent divergence in fingerprinting performance. Alpha-Z Divergence and Alpha Procrustes show the highest central tendencies across all granularity levels, approaching ceiling levels ($>$ 0.95) by 400–500 parcels, while their inter-quartile ranges contract, signalling consistently strong and stable identification. AI and Log-Euclidean distances track closely at coarse resolutions but plateau near 0.60 after 500 parcels, and their wider boxes and longer whiskers indicate greater variability once dimensionality increases. Bures–Wasserstein remains the least effective throughout, never surpassing the lower-mid accuracy band and showing the broadest dispersion of outcomes.

Collectively, the Fig. \ref{fig:parcellation_rank}(a) demonstrates two main trends. First, finer spatial resolution generally enhances individual identification, though gains diminish for most metrics beyond 600 parcels. Second, the relative ordering of methods becomes more pronounced with dimensionality, especially manifold-aware metrics Alpha-Z retains both high accuracy and low variance, whereas classical or noise-sensitive metrics (AI and Log-Euclidean) suffer noticeable performance degradation and spread. These patterns highlight the importance of choosing a robust distance measure when scaling analyses to high-resolution FCs.

Fig. \ref{fig:parcellation_rank}(b) presents a heat-map of median subject-identification accuracy (ID-rate) for five symmetric–positive-definite distance metrics as the cortical parcellation is progressively refined from 100 to 900 regions. Across all metrics, increasing the number of parcels generally boosts fingerprinting performance, but the magnitude of this gain differs markedly. Alpha-Z dominates at every granularity, climbing steeply from \(0.59\) at \(100\) parcels to \(>0.94\) by \(400\) parcels and plateauing at \(\approx0.96\) for \(600\) parcels and above. Alpha Procrustes shows the next-best performance, rising from \(0.54\) to \(0.65\) across the range, while AI and Log-Euclidean display more modest improvements (peaking around \(0.57\)). BW is the weakest throughout, never exceeding \(0.52\) and flattening near \(0.49\) for denser parcellations. These trends indicate that (i) finer parcellation enhances individual identification, but with diminishing returns beyond \(\approx600\) parcels, and (ii) the choice of metric is critical, with Alpha-Z offering a pronounced advantage over alternative SPD metrics.


The results clearly indicate that Alpha-Z Divergence and Alpha Procrustes Distance outperform geometry-aware metrics like AI Distance and Log Euclidean, particularly as the parcellation resolution increases. Alpha-Z Divergence emerges as the most robust metric, consistently yielding high identification rates across all tasks and resolutions, without the need for excessive regularization. In contrast, AI Distance and Log Euclidean Distance require substantial tuning and still fail to achieve comparable performance, especially for complex tasks like Emotion and Gambling. Figure \ref{fig:TASKS} also affirms that Alpha-Z Divergence achieves higher accuracy at all fMRI condition tasks (including Rest) than all other metrics  at fine resolutions. Finally, as granularity increases from 100 to 900 parcels, Alpha-Z and Alpha Procrustes maintain superior and stable task-specific identification rates, whereas AI, Log-Euclidean, and BW decline in accuracy and exhibit greater variability. The median identification-rate heatmap shows Alpha-Z climbing above 0.90 by 300 parcels (plateauing near 1.0) and a steadier ascent for Alpha Procrustes, while other manifold-based distances peak below 0.60, highlighting the scalability of geometry-aware Alpha-Z divergences.

\subsection*{Rank of SPD Matrices and Its Implications}

FC matrices are represented as symmetric correlation matrices, which are symmetric positive semidefinite \cite{bhatia2009positive}. The rank and invertibility of these matrices are directly related to their eigenvalues. Specifically, for an FC to be full-rank and invertible, all its eigenvalues must be strictly greater than zero. When one or more eigenvalues approach zero, the matrix becomes rank-deficient and non-invertible. This characteristic is particularly important as the number of brain regions (denoted as $m$) in the parcellation increases relative to the number of time points ($T$) in the BOLD signal.

The rank of an FC can be expressed as:
\[
\text{rank} \leq m \text{ for } T \geq m
\]
\[
\text{rank} < T \text{ for } T < m
\]

As parcellation resolution increases (i.e., as $m$ approaches or exceeds $T$), the resulting FC matrices are more likely to be rank-deficient. This creates challenges for traditional metrics like Log Euclidean and AI Distance, which rely on the invertibility of the matrices. The performance degradation observed for these metrics at higher resolutions can be attributed to the increasing rank-deficiency of the FC matrices, particularly when $m \geq T$.

\subsubsection*{Handling High-Dimensional FC Data}

Alpha-Z Divergence and Alpha Procrustes distance, by contrast, demonstrate robustness even in high-dimensional settings where FC matrices may approach a rank-deficient state as we can see in the Fig. \ref{fig:parcellation_rank}. These metrics do not rely as heavily on matrix invertibility \cite{dinh2021alpha,minh2022alpha,bhatia2009positive} and are less sensitive to the inherent challenges posed by high-dimensional parcellations. For instance, as shown in the analysis for 500–900 parcellations, Alpha-Z Divergence and Alpha Procrustes consistently deliver higher identification rates compared to AI Distance and Log Euclidean. This is particularly relevant for high-dimensional FC analysis, where the number of brain regions exceeds the number of time points in the data, a scenario where traditional metrics struggle.

\subsubsection*{Full-Rank Conditions}

In the preprocessing of the Destrieux parcellation, the FC matrices were generally full-rank\cite{abbas2021geodesic,bhatia2009positive} when the number of time points exceeded the number of regions. However, when the number of regions approached or exceeded the number of samples, rank deficiency became a concern for traditional metrics like AI Distance and Log Euclidean, further exacerbating their performance issues. In contrast, Alpha-Z Divergence and Alpha Procrustes demonstrated resilience, maintaining their high identification rates even when the matrices were approaching rank-deficiency. This indicates that these newer metrics are better suited for analyzing high-resolution parcellations where traditional metrics become unstable.

In summary, as parcellation resolution increases, the performance of traditional metrics such as AI Distance and Log Euclidean deteriorates due to rank-deficient matrices, particularly when the number of brain regions exceeds the number of samples. Alpha-Z Divergence and Alpha Procrustes, however, exhibit strong and consistent performance across all resolutions, making them highly suitable for high-dimensional FC analysis.

\subsection*{Regularization and Its Impact on Geodesic Distance}

\begin{figure}[H]
    \centering
    \includegraphics[width=\textwidth]{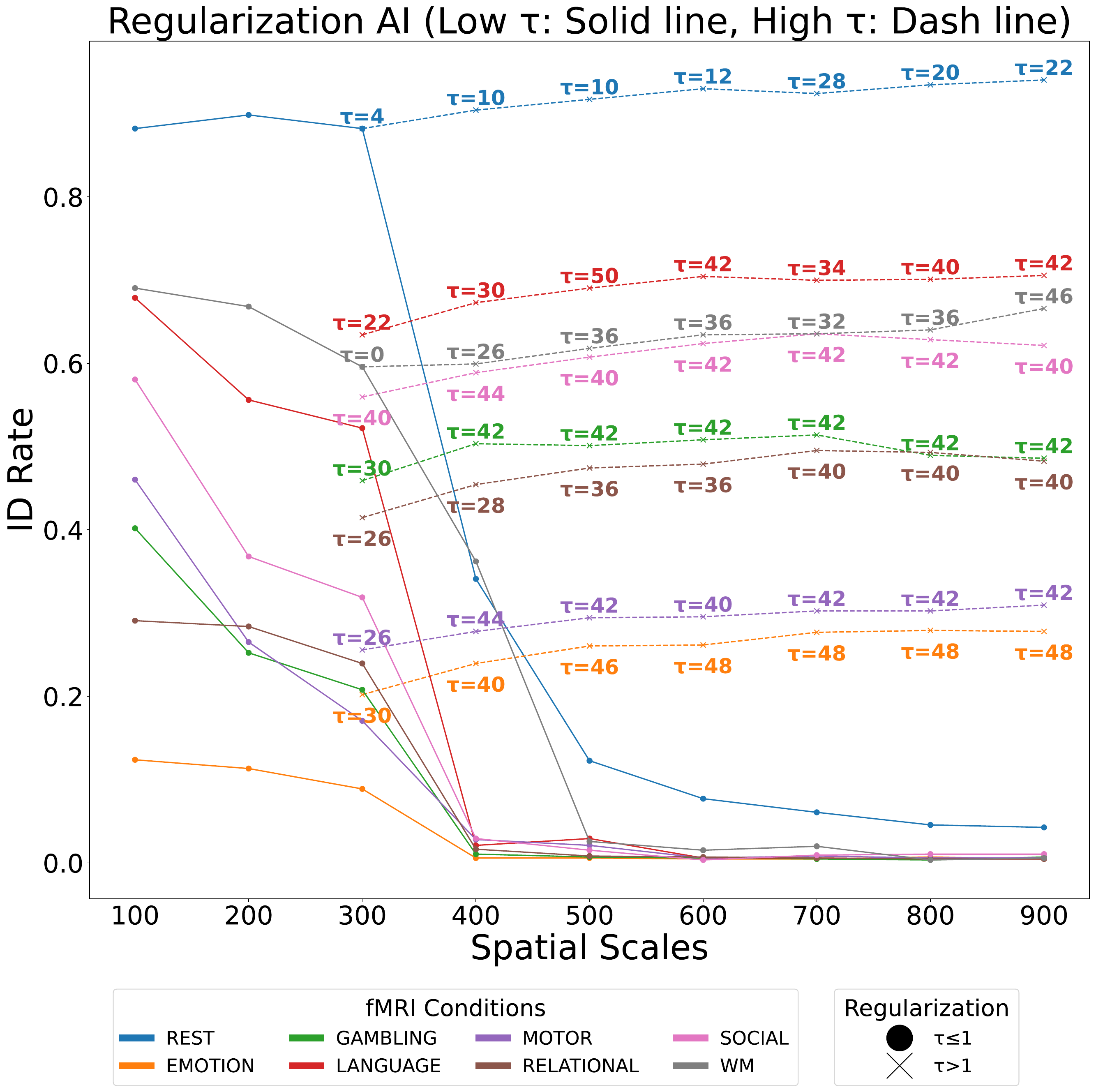}
    \caption{ Sensitivity of the Affine-Invariant (AI) distance metric to regularization strength across parcellation levels. The  solid line shows identification rates (ID rates) under low regularization conditions ($\tau \leq 1$), where performance rapidly declines as the number of parcels increases. The dash line illustrates the effect of higher regularization values ($\tau > 1$), where appropriate tuning improves performance stability across tasks and scales. These findings emphasize the critical role of regularization in geodesic-based metrics and motivate the need for more robust alternatives like which does not depend  on regularization parameter $\tau$ .
    }
    \label{fig:regularization}
\end{figure}

The impact of regularization\cite{venkatesh2020comparing,abbas2021geodesic} on geodesic distance performance was analyzed across different fMRI tasks, with the results presented in Fig.\ref{fig:regularization}. The figure illustrates the identification rate performance for AI comparison as a function of parcellation resolution, highlighting the role of the regularization parameter $\tau$ across a range of values ($\tau \leq 1$ and $\tau > 1$). The objective of this analysis was to assess how varying $\tau$ affects the identification rates across multiple parcellation resolutions (100–900 regions) and different fMRI tasks.

\subsubsection*{Effect of Low Regularization ($\tau \leq 1$)}

For low regularization values ($\tau \leq 1$) in solid line, the identification rate initially performs well at low parcellations (100–300 regions) but deteriorates rapidly as the number of parcellations increases. This effect is particularly pronounced for tasks like Rest, where the identification rate begins above 0.8 at 100 parcellations but declines to nearly zero by 500 parcellations. Similar trends are observed in tasks such as Emotion, Gambling, and Social, which also experience significant drops in performance as parcellation resolution increases. This pattern highlights the limitations of geodesic distance in handling high-dimensional FCs when low $\tau$ values are used, as the regularization is insufficient to maintain effective identification rates in these more complex settings.

\subsubsection*{Impact of High Regularization ($\tau > 1$)}

In contrast from the Fig.\ref{fig:regularization} in dash line, higher regularization values ($\tau > 1$) exhibit markedly different trends. For tasks with larger $\tau$ values, such as Gambling, Motor, and Social, identification rates remain more stable across all parcellation resolutions. For instance, in the Gambling task, identification rates stay above 0.5 even at 900 parcellations when $\tau = 42$, indicating that higher regularization parameters reduce the challenges posed by increased dimensionality.

The performance curves for regularization values ranging from $\tau = 10$ to $\tau = 48$ show that these higher values stabilize identification rates across most tasks, preventing the sharp declines seen in the low-$\tau$ settings. This stabilization is particularly beneficial for tasks like Social and WM, where the identification rates remain consistent across all parcellations. Furthermore, in tasks like Motor, high regularization values such as $\tau = 30$ and $\tau = 40$ result in relatively flat performance curves, maintaining identification rates around 0.6 even at higher parcellation resolutions.

The figure also reveals that the optimal regularization parameter $\tau$ varies across different tasks and parcellation resolutions. For example, the Rest condition experiences a sharp decline in identification rate performance at higher parcellations when $\tau \leq 1$, but this performance stabilizes when larger $\tau$ values are applied. Conversely, tasks like Language and Relational benefit more significantly from higher regularization values. These tasks maintain stable identification rates around 0.8 even at 900 parcellations when $\tau = 40$, illustrating the need for task-specific tuning of the regularization parameter to optimize performance.

Importantly, the results demonstrate that there is no fixed $\tau$ value that works optimally for all tasks and parcellations. For example, while Gambling and Social show significant improvements at higher $\tau$ values (e.g., $\tau = 42$), tasks such as Emotion and Rest benefit from more moderate increases in $\tau$. This variability underscores the importance of adjusting the regularization parameter based on both the specific task and the parcellation resolution in question.

This analysis highlights the critical role that regularization plays in optimizing geodesic distance performance for high-dimensional FC comparisons. While geodesic distance can perform reasonably well with low $\tau$ values for lower parcellation (100-300 resolutions) , the dimensional complexity of higher parcellations requires higher regularization values ($\tau = 10$ to $\tau = 48$) to maintain stable and accurate identification rates across tasks. Crucially, the optimal $\tau$ value is not fixed across tasks which is earlier said that tuning parameter is fixed\cite{abbas2021geodesic,abbas2023tangent,venkatesh2020comparing}; different tasks (e.g., Social, Gambling) benefit more from higher $\tau$ values, while others (e.g., Emotion, Rest) require more moderate regularization levels. This variability emphasizes the necessity of fine-tuning the regularization parameter according to the specific characteristics of the data being analyzed.

\subsection*{Alpha-Z Divergence preserves identifiability rankings of functional networks across spatial scales}

\begin{figure}[H]
    \centering
    \includegraphics[width=0.95\linewidth]{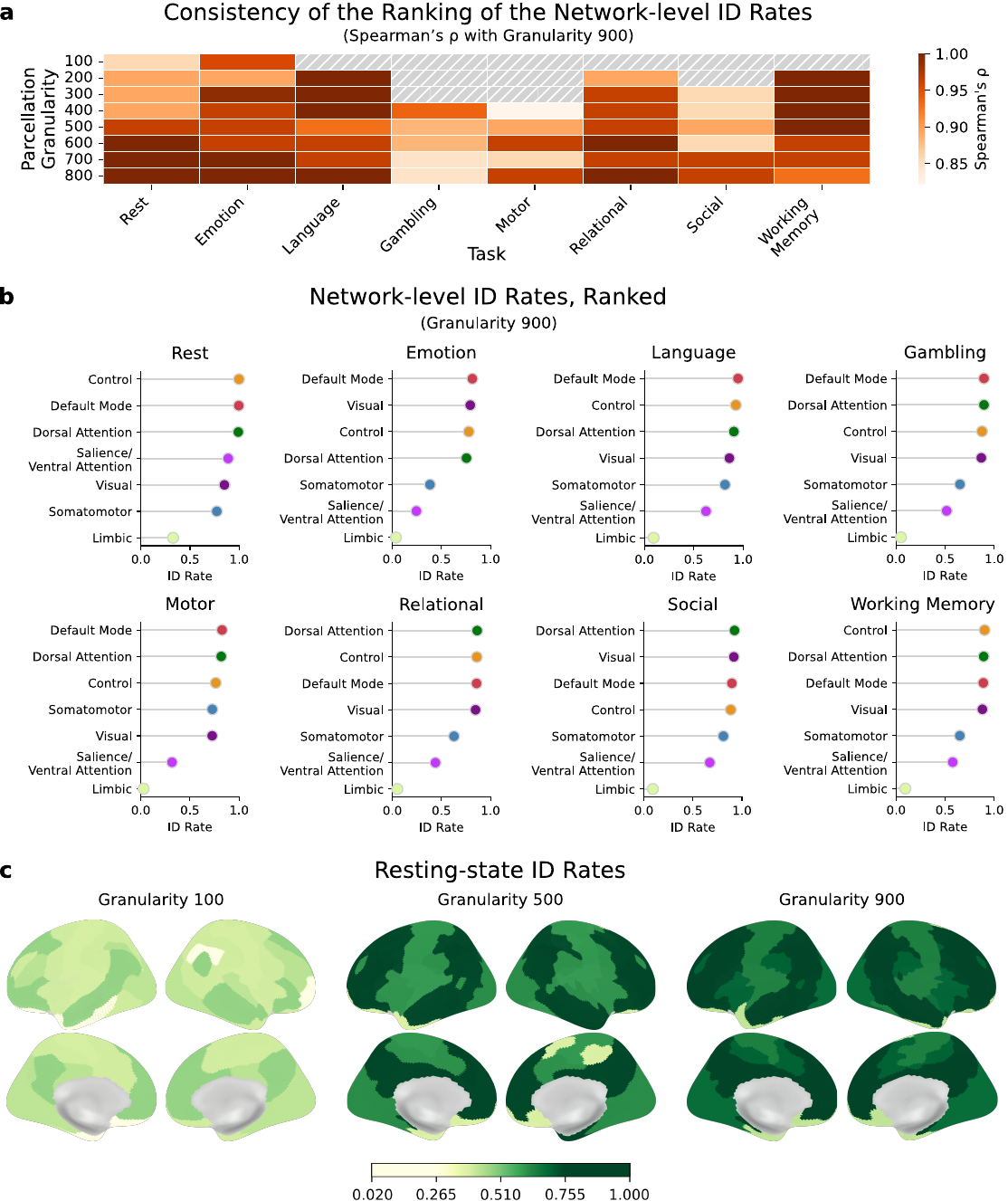}
    \caption{Using Alpha-Z Divergence, personal identification is feasible even when functional connectivity information is restricted to a single subset of the brain, \textit{i.e.}, a single Yeo resting-state network. When brain parcellation reaches a granularity of 400 or higher, the identifiability ranking of networks (based on identification rates) stabilizes across all HCP tasks and becomes consistent (via Spearman's rank-order correlation) with those observed at granularity 900 (panel \textbf{a}, gray cells indicate non-significant Spearman's correlation). By rendering the network rankings robust to spatial scaling, this stabilization strongly reflects the relationship between task structure and functional variability (panel \textbf{b}). For example, during resting state, the default mode and control networks exhibit higher identification rates than others, indicating greater group-level functional variability. This pattern persists across parcellation levels ranging from 100 to 900 (panel \textbf{c}), the consistency of which is reproducible across all tasks, resulting in other patterns outlined in \textbf{b}.}
    \label{fig:7}
\end{figure}

To test whether the alpha-Z divergence can reliably capture individual-specific signatures at the level of functional networks, we performed subject identification using only one of the seven Yeo networks \cite{yeo2011organization,nguyen2025evaluating}. This analysis was performed at varying levels of parcellation granularity (from 100 to 900 parcels) and at all seven HCP task states, plus the resting state. As expected, increasing parcellation granularity improved identification rates. Under the resting state condition, in particular, the default mode and frontoparietal control networks alone were sufficient for near-perfect identification at granularities 500 and above (figure \ref{fig:7}, b and c). This is in line with prior literature suggesting that transmodal networks, such as these two, carry rich information about personal functional traits \cite{finn2015functional,mantwill2022brain}, and these differences become increasingly accessible at finer spatial scales \cite{pena2018spatiotemporal}.

However, beyond absolute performance, we also asked whether the \textit{relative ranking} of network-level identification rates within a specific task remains stable across spatial scales. This question hinges on the idea that different task structures engage different functional systems, which should in turn exhibit varying levels of intersubject variability. For example, tasks that are tightly structured and rely on stereotyped perceptual processes (e.g., simple visual discrimination) may induce similar computations across subjects in visual regions, resulting in lower variability and, hence, lower identification rates. Conversely, tasks that elicit higher-order reasoning or flexible engagement of control systems should show greater interpersonal variability within relevant networks. If alpha-Z divergence is sensitive to such structure-function relationships, it should be able to capture these intersubject patterns even at spatial scales as coarse as 200 \cite{kong2023comparison}. This reasoning led to two possibilities. If network-level identification patterns consistent with task structure only emerge at fine granularity (e.g., 900), it would suggest that alpha-Z divergence requires high spatial resolution to detect task-relevant individual differences structure-function relationships. On the other hand, if such patterns are detectable and stable even at coarser scales, it would demonstrate that alpha-Z divergence is sensitive enough to pick up such relationships, despite the blurring effects introduced by coarser parcellation.

We found support for the latter. For each task, we computed Spearman’s rank-order correlation between the network-level identification rate rankings at coarser granularities and those at granularity 900. At this resolution, for all tasks except Emotion, the number of regions within each Yeo network (see table \ref{tab:numrois}) remained lower than the number of time points in the fMRI runs, ensuring that the corresponding FC matrices were not rank-deficient. We then tested whether these correlations exceeded chance expectations using permutation-based null models, with p-values corrected for the false discovery rate from multiple comparisons. This analysis revealed the granularity levels at which the network rankings began to converge with those at the finest scale (figure \ref{fig:7}a). Interestingly, for the resting state and the emotion task, these rankings stabilized as early as granularity 100, where the default mode and control networks already emerged as the most individually distinctive, and the somatomotor and limbic networks as the least. Other tasks required higher resolution: for instance, the motor task only began to exhibit stable rankings at granularity 400, with lower resolutions failing to achieve significant alignment with the rankings observed at 900.

In general, all HCP tasks exhibited stable network-level identification rate rankings starting at granularity 400 and above (figure \ref{fig:7}a). These stable rankings align well with expectations derived from task designs. For example, the Social task, which demands individualized processing of dynamic social cues, resulted in identification rates in the visual network that are among the highest, which suggests high variability in visual encoding strategies across individuals (figure \ref{fig:7}b). In contrast, the Language and Motor tasks, which present time-locked blocks of stimuli with predictable perceptual demands, elicited lower-ranked visual network identification rates, consistent with the suppression of intersubject variability in early sensory encoding \cite{liu2023individual} (figure \ref{fig:7}b).

Altogether, these results demonstrate that alpha-Z divergence not only robustly identifies individuals but also preserves meaningful signatures of task structure across a broad range of spatial scales. Its ability to uncover functionally relevant patterns, even at coarse granularities, suggests that it captures mesoscopic organizational features of human brain function. This expands the practical utility of connectome fingerprinting approaches in lower-resolution datasets and deepens our theoretical understanding of how individual traits are embedded within the fundamental, large-scale architecture of cognition.

\section*{DISCUSSION}

\subsection*{Superior Performance of Alpha-Z Divergence and Alpha Procrustes Distance}


This advantage becomes particularly pronounced as parcellation resolution increases and the complexity of the FCs grows, specifically in higher-dimensional settings where the rank of the SPD matrices and the effect of regularization become critical factors\cite{bhatia2009positive}. As discussed in the results section, increasing parcellation resolutions can lead to rank-deficiency in the SPD matrices, particularly when the number of brain regions ($m$) exceeds the number of samples ($T$). In such cases, geodesic distance like AI Distance, which rely on full-rank matrices for accurate computation, suffer significant performance degradation. In contrast, Alpha-Z Divergence and Alpha Procrustes Distance exhibit resilience to rank-deficiency and other high-dimensional challenges over other classical methods (Pearson and Euclidean), maintaining high identification rates across tasks, even as parcellations reach 900 regions, at the same time there tuning paprmeter is fixed. 
\subsubsection*{Regularization and Its Impact on AI Distance Performance}

The limitations of AI Distance were further highlighted in the analysis of regularization effects section. While regularization can help compensate for rank-deficiency\cite{amico2018,abbas2021geodesic}, AI Distance requires careful tuning of the regularization parameter $\tau$ to achieve optimal performance. As the results show, with low regularization values ($\tau \leq 1$), AI Distance experiences a steep decline in performance as parcellation resolution increases. Even with higher regularization values ($\tau > 1$), while performance stabilizes, AI Distance still cannot match the consistent and high identification rates of Alpha-Z Divergence and Alpha Procrustes Distance. This indicates that although AI Distance can benefit from regularization, it remains highly sensitive to the choice of $\tau$ and lacks the flexibility to perform well across a broad range of parcellation resolutions and tasks.

\subsubsection*{Stability and Flexibility of Alpha-Z Divergence and Alpha Procrustes Distance}

In contrast, Alpha-Z Divergence and Alpha Procrustes Distance ( with fixed tuning parameter {$\alpha=0.6$ for higher granularity}) demonstrate both stability and flexibility . They perform robustly across all tasks and parcellations without requiring extensive regularization, as evidenced by their high identification rates even in the absence of aggressive regularization. Alpha-Z Divergence while choosing fixed tuning parameter ( {$\alpha=0.99$} and {$\ z=1$}), in particular, achieves near-perfect identification rates across all parcellation resolutions, including 900 regions, indicating its ability to handle complex, high-dimensional FC data with minimal performance degradation. This robustness suggests that these metrics are well-suited for large-scale neuroimaging analyses, where the data are often high-dimensional and traditional metrics face difficulties in maintaining accuracy.

The analysis also emphasizes that the optimal regularization parameter $\tau$ for AI Distance is not constant across tasks or parcellation resolutions. Some tasks, like Gambling and Social, benefit more from higher $\tau$ values (e.g., $\tau = 42$), while others, such as Language and Rest, perform better with more moderate regularization (e.g., $\tau = 22$). This variability underscores the need for task-specific tuning when using AI Distance, adding complexity to its application in FC analysis. In contrast, Alpha-Z Divergence and Alpha Procrustes Distance demonstrate consistently high performance without the need for such extensive parameter tuning, making them more reliable and efficient for general application across different tasks and datasets.

Overall, the superior performance of Alpha-Z Divergence and Alpha Procrustes Distance is clear from their ability to handle high-dimensional FC data, maintain high identification rates across a wide range of tasks and parcellation resolutions, and function effectively without heavy reliance on regularization. These metrics offer a significant advantage over AI Distance, which requires fine-tuning of the regularization parameter and still struggles in higher-dimensional settings. Alpha-Z Divergence and Alpha Procrustes Distance’s ability to adapt to varying data characteristics while maintaining robust performance makes them ideal for future large-scale neuroimaging studies and FC analyses.

\subsection*{Advancement Over Traditional Metrics}

\begin{figure}[H]
    \centering
    \includegraphics[width=\textwidth]{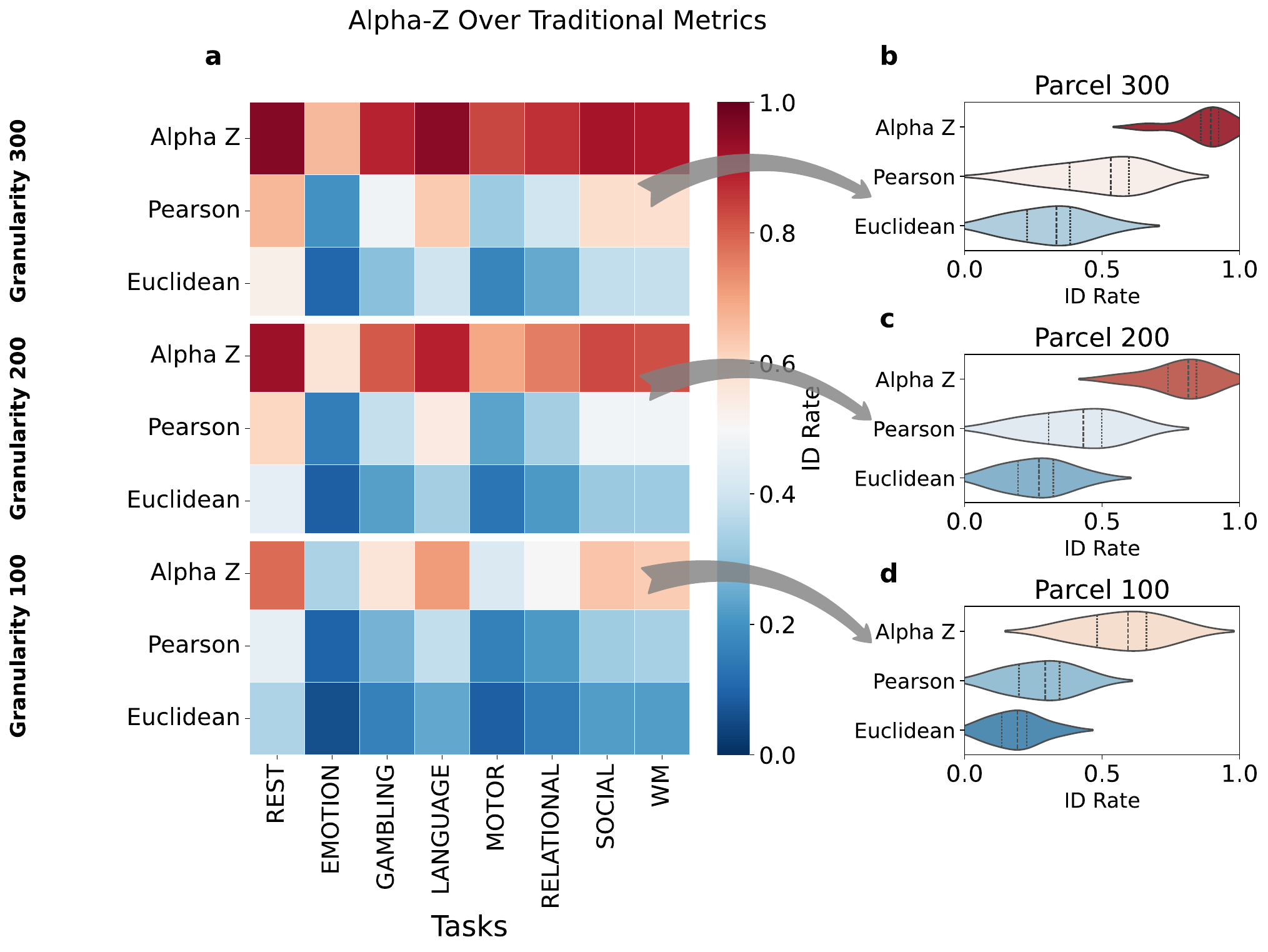}
    \caption{ Identification‐rate comparison between classical similarity measures and Alpha-Z Divergence.Panel \textbf{a} shows heat-maps of identification rates for Pearson correlation, Euclidean distance, and the proposed Alpha-Z Divergence at three parcellation levels (100, 200, 300 regions); warmer colours denote higher accuracy. Across every cognitive task, Alpha-Z yields markedly higher rates, illustrating its superior discriminative power even at low-to-moderate spatial resolutions. (Panels \textbf{(b-d}) depict violin plots of the same metrics at 300, 200, and 100 parcels, respectively, confirming that Alpha-Z maintains a consistently higher and less variable performance than the classical distances. Together, the results highlight the limitations of Pearson and Euclidean metrics and underscore the advantage of geometry-aware divergences for FC analysis.
    }
    \label{fig:advancement}
\end{figure}

The performance of Alpha-Z Divergence is clearly shown to be superior when compared to traditional metrics \cite{venkatesh2020comparing} like Pearson correlation and Euclidean distance \cite{bravais1846, galton1886}, as depicted in Fig.\ref{fig:advancement}. This advancement becomes more pronounced as the number of parcellations increases, highlighting the robustness of Alpha-Z Divergence in identifying FCs across a wide range of tasks. The figure presents a comparison of identification rates across three different parcellation resolutions: 100, 200, and 300 parcellations, providing a clear view of how each metric performs in different settings.

In the (Fig. \ref{fig:advancement}a)(Granularity 100) and (panel \textbf{d}), Alpha-Z Divergence consistently outperforms both Pearson correlation and Euclidean distance across all tasks. This superiority is particularly evident in tasks like Rest, where Alpha-Z Divergence achieves an identification rate of above 0.8, compared to Pearson’s 0.5 and Euclidean’s 0.4. The gap between Alpha-Z Divergence and traditional metrics is substantial, especially in tasks such as Emotion and Motor, where Pearson and Euclidean struggle to maintain accuracy. These results indicate that Alpha-Z Divergence is more capable of capturing the nuanced relationships within the FC matrices, even at relatively low resolutions compared to traditional metrics.

As granularity increase to 200 the (Fig.\ref{fig:advancement}a)) shows a continuation of the trend, with Alpha-Z Divergence maintaining its superior performance. While Pearson correlation shows slight improvements across some tasks, such as Gambling and Relational, it still lags behind Alpha-Z Divergence, particularly in tasks like Rest and Social. It is also clear from the (panel \textbf{c}) that Euclidean distance continues to show weaker performance, with identification rates remaining around 0.4 for most tasks. Alpha-Z Divergence, on the other hand, maintains high accuracy, consistently achieving identification rates above 0.9 across nearly all tasks. This further highlights the ability of traditional metrics to handle increasing parcellation complexity without sacrificing identification accuracy.

As parcellation resolution increases to 300 regions (Fig.\ref{fig:advancement}b), the gap between Alpha-Z Divergence and traditional metrics widens even further. (Fig.\ref{fig:advancement}a) shows  Pearson correlation and Euclidean distance experience significant performance drops in tasks such as Language, Motor, and Social, where identification rates fall below 0.5 for both metrics. Conversely, Alpha-Z Divergence maintains its high identification rates, reaching almost 1.0 for tasks like Rest and Language. This result showcases the robustness of Alpha-Z Divergence in high-dimensional settings, where traditional metrics fail to fully capture the complex structure of high-resolution FCs.

Overall, the results presented in Fig. \ref{fig:advancement} demonstrate the clear advancement of Alpha-Z Divergence over traditional metrics like Pearson correlation and Euclidean distance. Across all parcellation resolutions and tasks, Alpha-Z Divergence consistently delivers higher identification rates, showcasing its superior capability in capturing the complex and intricate relationships within FC data. This is particularly important as the dimensionality of the data increases, where traditional metrics exhibit significant performance degradation.The strong and stable performance of Alpha-Z Divergence again proof that it is an ideal candidate for use in high-resolution FC analysis, where accuracy and robustness are paramount.

\subsection*{Eigenvalue Information and Its Impact on Matrix Rank}
The eigenvalue distribution is a crucial factor in understanding the performance of various distance metrics and their ability to handle FC data, particularly as the dimensionality of the data increases.The behavior observed in the eigenvalue curves emphasizes that with increasing parcellation resolution, the matrices are becoming increasingly ill-conditioned. This ill-conditioning arises due to the accumulation of eigenvalues close to zero, which makes it more difficult for distance metrics to accurately compare matrices without sufficient regularization.

The rank-deficiency, as indicated by the increasing proportion of near-zero eigenvalues, explains why metrics such as AI Distance and Log Euclidean, which depend on the matrix being full-rank, suffer significant performance degradation at higher parcellations. In contrast, metrics like Alpha-Z Divergence and Alpha Procrustes Distance are less sensitive to this issue and continue to perform well despite the increasing number of small eigenvalues. These metrics are designed to handle the intrinsic geometric structure of the FCs more robustly, even when the data become rank-deficient.


The growing proportion of eigenvalues close to zero at higher parcellations has a direct impact on the ability of distance metrics to function effectively. Metrics that rely on matrix invertibility or assume full-rank matrices struggle as the eigenvalue distribution becomes skewed towards zero. This is where Alpha-Z Divergence and Alpha Procrustes Distance show their advantage. Unlike traditional metrics, these newer metrics do not depend on full-rank matrices and can effectively handle the increased dimensionality of the data without suffering the same performance declines.


The eigenvalue highlights the critical role that matrix rank plays in the performance of distance metrics for FC comparisons. As parcellation resolution increases, the growing number of eigenvalues close to zero leads to rank-deficiency, which impairs the effectiveness of traditional metrics like AI Distance. The robustness of Alpha-Z Divergence and Alpha Procrustes Distance, even in the presence of a large number of near-zero eigenvalues, underscores their suitability for high-dimensional FC analysis, making them the metrics of choice for scenarios where matrix rank is compromised.

\subsection*{3D Visualization of Performance: AI Distance vs. Alpha-Z Divergence}

\begin{figure}[H]
    \centering
    \includegraphics[width=\textwidth]{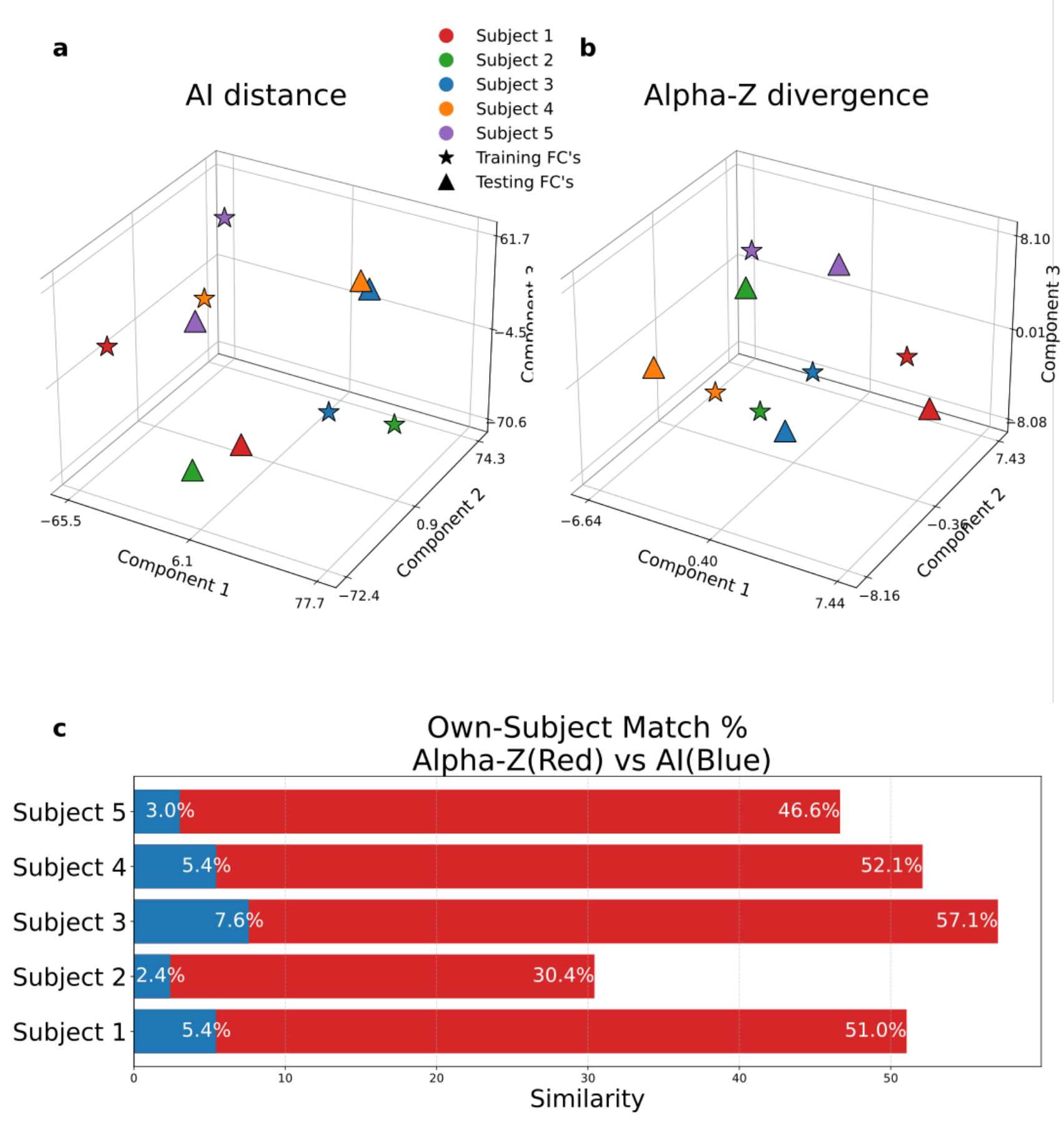}
    \caption{3D visualization of subject-level embedding using two different distance measures: Affine-Invariant (AI) distance (panel \textbf{a}) and the proposed Alpha-Z divergence (panel \textbf{b}). Each point represents a subject’s left-right fMRI pair projected in reduced component space. The percentages in (panel \textbf{c}) indicate  accurately labeled participants for these five random participants chosen from Rest task (400 spatial). While the AI metric fails to clearly cluster corresponding pairs, Alpha-Z achieves significantly  better alignment and separation, resulting in improved identification accuracy.
    }
    \label{fig:ai_vs_alpha}
\end{figure}

Figures \ref{fig:ai_vs_alpha}a and \ref{fig:ai_vs_alpha}b provide a 3D visualization\cite{venkatesh2020comparing} comparing the performance of AI Distance and Alpha-Z Divergence in clustering FCs from the same subject across different sessions. These visualizations are based on data from the Rest task at 400 parcellations, plotted across three principal components, with the percentage of same-subject (five subjects chosen randomly from the 428 subjects) labeled annotated for each subject.

In Fig.~\ref{fig:ai_vs_alpha}a, which depicts results for the Affine-Invariant (AI) distance, clustering of FCs from the same subject shows large variation in same-subject matching. For example, panel \ref{fig:ai_vs_alpha}c reports correct-match rates of 7.6\% and 5.4 \% for Subjects 3 and 4, whereas Subject 2 achieves only 2.4 \%. This variability indicates that AI distance struggles to consistently group FCs from the same individual, yielding a scattered and less coherent distribution of points in the 3-D embedding.

Moreover, the visualization reveals significant overlap between FCs from different subjects. Notably from  Fig.(\ref{fig:ai_vs_alpha}c) for AI, subjects such as $1$, $3$, and $4$ show higher mis-labeling rate that are clustered closely with points from other individuals, indicating that AI Distance has difficulty distinguishing between subjects. The considerable overlap and broad distribution of points suggest that AI fails to capture the unique structural similarities within FCs of the same individual across different sessions, particularly at higher parcellation levels. 

In contrast, Fig.(\ref{fig:ai_vs_alpha}b), which depicts the performance of Alpha-Z Divergence, demonstrates markedly improved clustering of FCs for the same subject. In the Fig. (\ref{fig:ai_vs_alpha}c) for Alpha-Z, the same-subject correct-match percentages are significantly higher across all participants. For instance, subject $1$ achieves a similarity rate of 51.0\%, a dramatic improvement over the results achieved with AI distance. Similarly, subjects $2$ and $5$, which exhibited lower labeling rates using AI, show much better performance with Alpha-Z Divergence, achieve correct-match rates of 30.4\% and 46.6\%, respectively.

The clustering of data points is far more distinct and tighter with Alpha-Z Divergence, indicating that it better captures the intrinsic similarities within the FCs of the same individual. Importantly, there is minimal overlap between FCs from different subjects, suggesting that Alpha-Z Divergence is much more effective at differentiating between individuals. The clear separation of points in the 3D embedding further confirms that Alpha-Z Divergence excels at preserving subject-specific information, even in the context of high-dimensional data. 

Figure \ref{fig:ai_vs_alpha} provide compelling evidence of the superior performance of Alpha-Z Divergence over AI Distance in identifying FCs from the same subject. AI  exhibits substantial overlap between FCs from different subjects and generally lower same-subject matching rates, reflecting its limitations in capturing FC structure across sessions at high parcellations. In contrast, Alpha-Z Divergence shows tighter clustering, minimal overlap, and consistently higher same-subject identification percentages. These results underline Alpha-Z Divergence’s robustness and efficacy in handling high-dimensional FC data, making it a more suitable metric for FC analysis, particularly in scenarios where AI Distance struggles.

\subsection*{Validation of Performance via Null Model Analysis}

To establish the statistical significance of the identification performance achieved with the Alpha-Z divergence, we performed a comprehensive null model analysis across eight cognitive tasks and multiple spatial parcellations (ranging from 100 to 900 regions).

\begin{figure}[H]
    \centering
    \includegraphics[width=\textwidth]{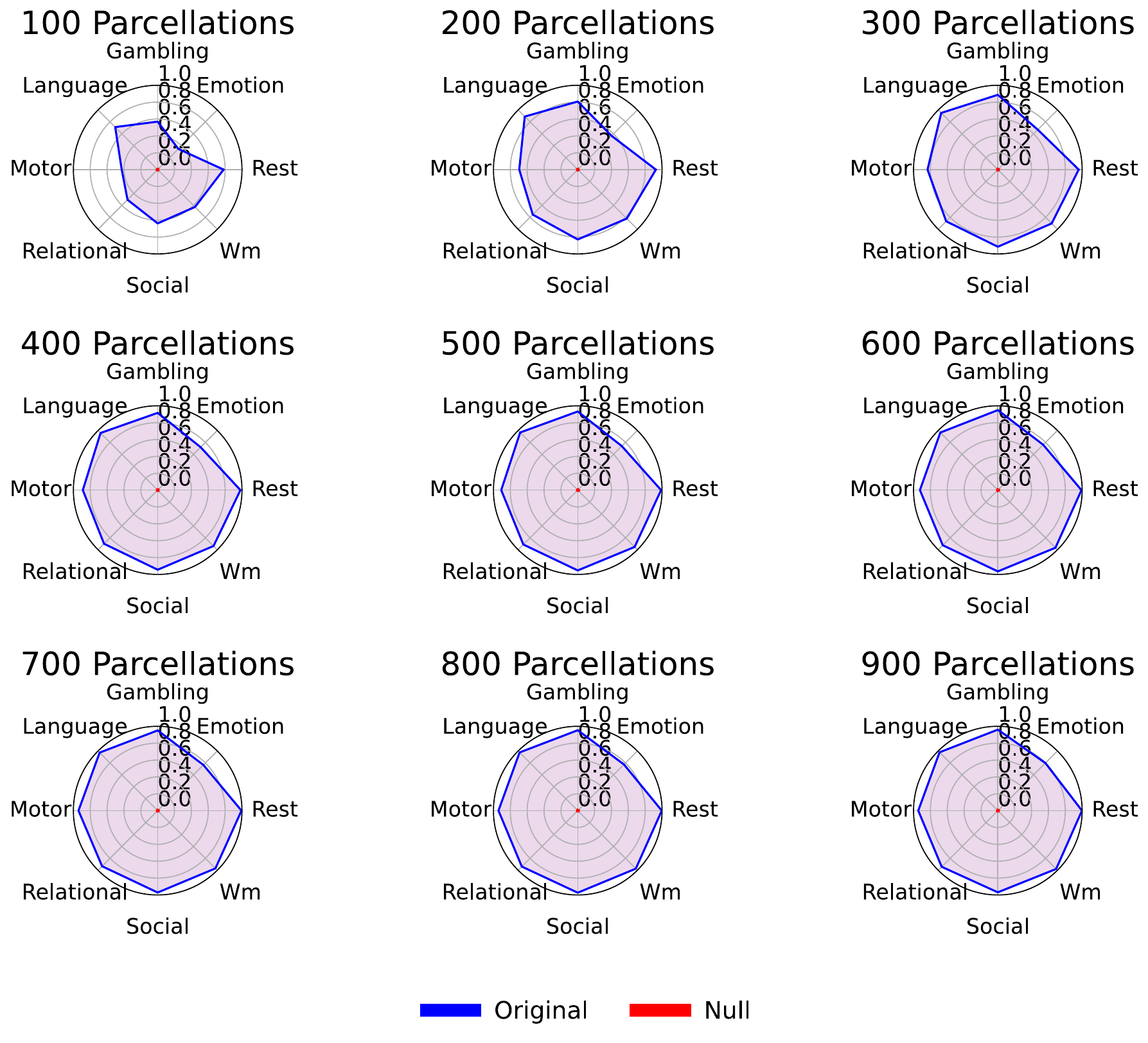}
    \caption{Null model analysis validating the statistical significance of observed identification rates across tasks and parcellation levels. Each radar chart compares the performance of the original data (blue) against a null model (red) across eight cognitive tasks for a specific parcellation resolution (ranging from 100 to 900). The consistent separation between original and null performances across all settings confirms that the observed results are not due to chance and reflect meaningful subject-level discriminability driven by the underlying brain connectivity patterns.
    }   
\label{fig:null_model}
\end{figure}

As illustrated in Fig.\ref{fig:null_model}, each radar plot compares the identification accuracy from the original data (blue) to that of a null model (red), where subject labels were randomly permuted to disrupt individual-specific signal. 

Across all tasks and parcellation levels, the null model consistently yielded identification rates close to chance level approximately 10\%, whereas the original data maintained substantially higher accuracy aproximately 95\%  for other taks and 100\% for rest task. This consistent separation confirms that the performance of Alpha-Z divergence is not driven by random effects or label artifacts, but instead captures meaningful subject-specific brain connectivity structure. These results provide strong evidence for the robustness and discriminative power of Alpha-Z divergence in FC analysis.

\subsection*{Limitations of the study}

While our findings underscore the robustness of the Alpha-Z divergence across high-dimensional FCs, at the same time it has several limitations. First, our assessment was conducted using high-quality data from the Human Connectome Project (HCP); thus, the generalization of performance to datasets with lower signal-to-noise ratios or clinical populations (e.g., elderly or patient cohorts) remains uncertain and necessitates targeted validation. Lastly, while identification rates show strong discriminative capability, the specific neurobiological substrates driving variations in Alpha-Z distances are not yet fully understood, highlighting the need for future investigations that associate metric behavior with underlying structural and physiological features.

\section*{ACKNOWLEDGMENTS}


This work was funded by the National Science Foundation via grant NO. 2153492. The authors thank all members of the lab for their support.

\bibliography{references}



\newpage



\section*{SUPPLEMENTARY MATERIALS: METHODS}


\subsection*{KEY RESOURCES TABLE}
\begin{table}[htbp] 
\centering
\renewcommand{\arraystretch}{1.4}
\begin{tabular}{|p{5cm}|p{4cm}|p{6cm}|}
\hline
\textbf{ RESOURCE} & \textbf{SOURCE} & \textbf{IDENTIFIER} \\
\hline
\multicolumn{3}{|l|}{\textbf{Deposited data}} \\
\hline
Time series and functional connectomes of HCP dataset & This paper & \url{https://auburn.app.box.com/folder/244365587358} \\
\hline
Functional connectomes of Validation dataset & This paper & \url{https://auburn.app.box.com/folder/244365587358} \\
\hline
\multicolumn{3}{|l|}{\textbf{Software and algorithms}} \\
\hline
PYTHON & PYTHON & PYTHON \\
\hline
Packages & spd-metrics-id & \url{https://pypi.org/project/spd-metrics-id/}\\
\hline
Freesurfer & Laboratory for Computational Neuroimaging at the Athinoula A. Martinos Center for Biomedical Imaging & \url{https://surfer.nmr.mgh.harvard.edu/} \\
\hline
AFNI & National Institute of Health & \url{https://afni.nimh.nih.gov/} \\
\hline
FSL & Analysis Group, FMRIB, Oxford, UK & \url{https://fsl.fmrib.ox.ac.uk/fsl/fslwiki} \\
\hline
\end{tabular}
\label{tab:key_resources}
\end{table}

\subsection*{RESOURCE AVAILABILITY }
\subsubsection*{Lead contact}


Requests for further information and resources should be directed to and will be fulfilled by the lead contact, Kaosar Uddin (mzu0014@auburn.edu).

\subsubsection*{Resource Availability}


Datasets generated in this study have been deposited to [FC of Validation dataset, \& \url{https://auburn.app.box.com/folder/244013697386}].

\subsubsection*{Data and code availability}


\begin{itemize}
    \item The Human Connectome Project data can be acquired from \href{https://www.humanconnectome.org/}{Connectome HCP Young Adult data homepage}.
    \url{https://www.humanconnectome.org/}. 
    \item All original code has been deposited at Github \url{https://github.com/KaosarUddin/b_f} and \url{https://hub.docker.com/r/kaosar148/spd-metrics-id} is publicly available  \cite{uddin2025spdmetricsid}.
    \item Any additional information required to reanalyze the data reported in this paper is available from the lead contact upon request.    
\end{itemize}

\subsection*{EXPERIMENTAL MODEL AND STUDY PARTICIPANT DETAILS}



\subsection*{Experimental data}


In this study, we utilized a set of functional brain atlases, specifically the Schaefer parcellation of the cortex. This parcellation is derived from resting-state fMRI data collected from 1,489 participants, which were aligned using surface-based registration techniques. To generate the Schaefer parcellation, a gradient-weighted Markov random field approach was employed, combining local gradient information with global similarity metrics. The Schaefer parcellation is available in ten levels of granularity, ranging from 100 to 1000 regions in increments of 100. These parcellations are provided in both volumetric and grayordinate formats. Since the grayordinate parcellations share the same surface space as the HCP fMRI data, they can be mapped onto the fMRI data with relative ease. Surface-based mapping offers superior alignment between the fMRI data and the Schaefer parcellations compared to volumetric mapping. Therefore, we used surface-based mapping to align the 100–900 region Schaefer parcellations with the fMRI data. During the data processing phase of this study, we were unable to successfully map the 1,000 region Schaefer parcellation for the HCP Young Adult dataset. Additionally, 14 subcortical regions were integrated into each parcellation, as provided by the HCP release \texttt{filename: Atlas\_ROI2.nii.gz}.This file was converted from NIFTI to CIFTI format using the HCP Workbench software\href{https://www.humanconnectome.org/software/connectome-workbench.html}{HCP Workbench software}, via the command \texttt{wb\_command -cifti-create-label}. For example, the Schaefer-100 parcellation resulted in a total of 114 brain regions, and the Schaefer-900 parcellation resulted in a total of 914 brain regions. The Schaefer parcellation atlases contain labels of Yeo canonical functional networks \cite{yeo2011organization,schaefer2018local} whose numbers of regions for all parcellation levels are included in Table \ref{tab:numrois}.

\begin{table}[ht]
\caption{Number of ROIs per Yeo network across Schaefer parcellation granularities.}
\label{tab:network_counts}
\begin{tabular}{lrrrrrrr}
\toprule
 & Visual & Somatomotor & \makecell{Dorsal \\ Attention} & \makecell{Salience / \\ Ventral \\ Attention} & Limbic & Control & \makecell{Default \\ Mode} \\
Granularity &  &  &  &  &  &  &  \\
\midrule
100 & 17 & 14 & 15 & 12 & 5 & 13 & 24 \\
200 & 29 & 35 & 26 & 22 & 12 & 30 & 46 \\
300 & 47 & 57 & 34 & 34 & 20 & 40 & 68 \\
400 & 61 & 77 & 46 & 47 & 26 & 52 & 91 \\
500 & 74 & 96 & 56 & 59 & 33 & 69 & 113 \\
600 & 89 & 112 & 72 & 73 & 42 & 82 & 130 \\
700 & 120 & 128 & 87 & 75 & 49 & 92 & 149 \\
800 & 134 & 151 & 99 & 87 & 54 & 105 & 170 \\
900 & 147 & 173 & 104 & 105 & 60 & 117 & 194 \\
\bottomrule
\end{tabular}
\label{tab:numrois}
\end{table}

In this work, we used data from the HCP 1,200 participants release \cite{van2013wu} and extracted three different subsets. The first consists of 428 unrelated participants (223 women, mean age: 28.67 years old, range: 22–36) selected. 

\paragraph{Preprocessing of HCP Dataset:}

The HCP dataset underwent a "minimal" preprocessing pipeline, which included artifact removal, motion correction, and registration to a standard template, as detailed in earlier publications. To further process the resting-state fMRI data, we added the following steps: (i) regressed out the global gray matter signal from voxel time courses, (ii) applied a first-order Butterworth bandpass filter in both forward and reverse directions [0.001–0.08Hz; MATLAB functions butter and filtfilt], and (iii) z-scored and averaged voxel time courses for each brain region, excluding outlier time points beyond three standard deviations from the mean \texttt{(workbench software, wb\_command -cifti-parcellate)}. 

FC matrices were constructed by computing the Pearson correlation coefficient between the mean time series of every pair of brain regions. This resulted in symmetric, weighted adjacency matrices with values ranging from –1 to 1. FC matrices were computed for each participant individually.

\paragraph{Preprocessing of Validation Dataset:}
The validation dataset was processed using an in-house pipeline based on AFNI, FSL, and MATLAB, adhering to state-of-the-art guidelines. The same cortical parcellation scheme (Schaefer parcellation) introduced earlier was employed, while subcortical regions were derived from the Tian parcellation at scale I.

\subsection*{Distance Description}\label{sec2}

\subsubsection*{1. Affine Invariant (AI) Distance}

The Affine Invariant (AI) distance is a robust measure used to compare covariance matrices, particularly in the context of diffusion tensor imaging (DTI) and brain connectivity analysis. This distance metric is invariant under affine transformations, making it especially useful when the data undergoes non-linear deformations. The AI distance between two positive definite matrices $A$ and $B$ is defined as:

\[
d_{\text{AI}}(A,B) = \| \log(A^{-1/2} B A^{-1/2}) \|_F
\]

where $\log$ denotes the matrix logarithm and $\|\cdot\|_F$ represents the Frobenius norm. This distance captures the dissimilarity between matrices by accounting for both shape and orientation, making it particularly useful for tasks that involve structural variability.

\subsubsection*{2. Log-Euclidean Distance}

The Log-Euclidean distance is a metric designed to measure the distance between symmetric positive definite (SPD) matrices by leveraging the Log-Euclidean framework. This method simplifies the comparison of SPD matrices by applying the matrix logarithm to transform the original space into a Euclidean space. The distance between two SPD matrices $A$ and $B$ is given by:

\[
d_{\text{LE}}(A,B) = \| \log(A) - \log(B) \|_F
\]

where $\log(\cdot)$ is the matrix logarithm and $\|\cdot\|_F$ is the Frobenius norm. The Log-Euclidean distance is advantageous due to its computational efficiency and the ability to retain the geometric properties of the space, making it suitable for various applications in brain imaging and functional connectivity.

\subsubsection*{3. Bures-Wasserstein (BW) Distance}

The Bures-Wasserstein (BW) distance is a metric that stems from optimal transport theory, specifically tailored for comparing probability measures with a focus on Gaussian distributions. In the context of covariance matrices, the BW distance between two SPD matrices $A$ and $B$ can be expressed as:

\[
d_{\text{BW}}(A,B) =\left(\text{tr}(A) + \text{tr}(B)-2\text{tr}\left((A^{1/2} B A^{1/2})^{1/2}\right) \right)^{1/2}
\]

where $\text{tr}(\cdot)$ denotes the trace of a matrix. The BW distance captures both the spread (variance) and the mean (location) of the distributions, making it a powerful tool for comparing functional connectomes and other brain imaging data where Gaussian assumptions are reasonable.

\subsubsection*{4. Alpha-Procrustes Distance}

The Alpha-Procrustes distance defines a parametrized family of metrics on the space of symmetric positive definite (SPD) matrices, generalizing both the Bures-Wasserstein and Log-Euclidean distances. This distance emerges from an extension of the Procrustes distance problem, which aims to align two shapes as closely as possible under a set of transformations.

The Alpha-Procrustes distance between two SPD matrices $A$ and $B$ is defined as:

\[
d_{\alpha}^{\text{Pro}}(A,B) = \min_{U \in U(n)} \|A^{\alpha} - B^{\alpha}U\|_F
\]

where $U(n)$ denotes the set of unitary matrices of size $n \times n$, and $\|\cdot\|_F$ is the Frobenius norm. The parameter $\alpha$ modulates the influence of the transformation, with specific values of $\alpha$ corresponding to well-known distances:

\begin{itemize}
    \item $\alpha = \frac{1}{2}$: This case corresponds to the Bures-Wasserstein distance, scaled to match its conventional form.
    \item $\alpha = 0$: This case results in the Log-Euclidean distance, which reflects the Riemannian distance in the space of SPD matrices.
\end{itemize}

\paragraph*{Riemannian Geometry Interpretation}

The Alpha-Procrustes distance can also be interpreted as the Riemannian distance associated with a family of Riemannian metrics on the manifold of SPD matrices. This family encapsulates both the Log-Euclidean and Wasserstein Riemannian metrics as special cases, thereby offering a unified framework for these distances.

In special Cases, Alpha Procrustes distance introduces  Bures-Wasserstein Distance and Log-Euclidean Distance, 

\paragraph{Bures-Wasserstein Distance ($\alpha = \frac{1}{2}$):}

\[
d_{\frac{1}{2}}^{\text{Pro}}(A,B) = 2 \left( \text{tr}\left(A + B - 2\left(A^{1/2}BA^{1/2}\right)^{1/2}\right) \right)^{1/2}
\]

\paragraph{Log-Euclidean Distance ($\alpha \rightarrow 0$):}

\[
\lim_{\alpha \to 0} d_{\alpha}^{\text{Pro}}(A,B) = \| \log(A) - \log(B) \|_F
\]

\paragraph*{Generalization to Infinite-Dimensional Spaces}

The concept of Alpha-Procrustes distance is further extended to positive definite Hilbert-Schmidt operators on an infinite-dimensional Hilbert space. This extension includes the Bures-Wasserstein and Log-Hilbert-Schmidt distances, making it applicable in settings such as Gaussian measures and reproducing kernel Hilbert spaces (RKHS).

The Alpha-Procrustes distance is particularly effective in scenarios where the comparison of SPD matrices is required, such as in brain connectivity analysis, diffusion tensor imaging, and shape analysis. It offers a versatile and mathematically robust framework for modeling dissimilarities between SPD matrices under various transformations and metrics.

\begin{table}[htbp] 
\begin{adjustbox}{width=\textwidth,center}
\centering
\renewcommand{\arraystretch}{1}
\begin{tabular}{>{\raggedright\arraybackslash}p{2.5cm} >{\raggedright\arraybackslash}p{4cm} c c c}
\toprule
\textbf{Metric} & \textbf{Formula} & \textbf{Geodesic} & \textbf{Performance} & \textbf{Tuning Sensitivity} \\
\midrule
Affine-Invariant & $\left\| \log(X^{-1/2} Y X^{-1/2}) \right\|_F$ & Yes & Needs tuning & High \\
Log-Euclidean & $\left\| \log(X) - \log(Y) \right\|_F$ & Yes & Needs tuning & High \\
Bures-Wasserstein & \makecell{
$\mathrm{tr}(X) + \mathrm{tr}(Y) - 2\,\mathrm{tr}\left(\left(X^{1/2} Y X^{1/2}\right)^{1/2}\right)$
} & Yes & No need & No \\

Alpha-Procrustes & $\min_{U \in \mathcal{U}(n)} \left\| X^\alpha - Y^\alpha U \right\|_F$ & Yes* & Good & Medium** \\
Alpha-z-Bures Wasserstein & $\mathrm{tr}((1 - \alpha)X + \alpha Y) - \mathrm{tr}(Q_{\alpha,z}(X,Y))$ & No & Best & Low*** \\
Euclidean & $\left\| X - Y \right\|_F$ & No & Needs tuning & High \\
Pearson Correlation & $\displaystyle 1 - \frac{\mathrm{cov}(X, Y)}{\displaystyle \sigma_x \, \sigma_y}$
 & 
No & Needs tuning & High \\
\bottomrule
\end{tabular}
\end{adjustbox}
\caption{Performance-based comparison of distance metrics across parcellation scales.}
\captionsetup{justification=justified,singlelinecheck=false}
\caption*{\small $^*$ In special cases Alpha Procrust is a geodesic distance ; $^{**}$ Medium but the tuning parameter is fixed for higher parcellations; $^{***}$ Low and the best thing is the tuning parameter is fixed.}
\label{tab:performance_comparison}
\end{table}

\subsubsection*{5.Alpha-z-Bures Wasserstein Divergence}

The Alpha-z Divergence, as described in the paper, introduces a new divergence measure specifically tailored for positive semidefinite matrices. This divergence is referred to as the Alpha-z-Bures Wasserstein Divergence and serves as a generalization of the classical Bures-Wasserstein distance. The divergence is defined mathematically as:

\[
\Phi(A,B) = \text{Tr}((1 - \alpha)A + \alpha B) - \text{Tr}(Q_{\alpha,z}(A,B))
\]

where $A$ and $B$ are positive semidefinite matrices, and $Q_{\alpha,z}(A, B)$ is defined as:

\[
Q_{\alpha,z}(A, B) = \left(A^{\frac{1-\alpha}{2z}} B^{\frac{\alpha}{z}} A^{\frac{1-\alpha}{2z}}\right)^z
\]

This matrix function $Q_{\alpha,z}(A, B)$ is derived from the alpha-z-Rényi relative entropy, which is a family of entropy measures used in quantum information theory.

\paragraph*{Key Properties of Alpha-z-Bures Wasserstein Divergence}

\begin{enumerate}
    \item \textbf{Quantum Divergence:} The Alpha-z-Bures Wasserstein divergence is shown to be a quantum divergence, satisfying several essential properties, such as:
    \begin{itemize}
        \item \textbf{Non-negativity:} $\Phi(A, B) \geq 0$ with equality if and only if $A = B$.
        \item \textbf{Data Processing Inequality:} This divergence is invariant under completely positive trace-preserving maps, which implies that it satisfies the data processing inequality in quantum information theory.
    \end{itemize}
    
    \item \textbf{In-Betweenness Property:} The divergence also satisfies the in-betweenness property, meaning that for any pair of positive semidefinite matrices $A$ and $B$, and any matrix power mean $\mu_p(t; A, B)$ with $p \in [1/2, 1]$, the inequality $\Phi(A, \mu_p(t; A, B)) \leq \Phi(A, B)$ holds. This property ensures that the divergence between $A$ and the power mean $\mu_p(t; A, B)$ is always less than or equal to the divergence between $A$ and $B$.
\end{enumerate}

The Alpha-z-Bures Wasserstein divergence is particularly useful in quantum information theory, where it can be applied to measure the dissimilarity between quantum states represented by positive semidefinite matrices. Its ability to generalize well-known divergences and its compatibility with quantum mechanical operations make it a versatile tool for analyzing quantum systems.

The introduction of the Alpha-z-Bures Wasserstein divergence provides a new and flexible framework for quantifying differences between positive semidefinite matrices, extending traditional concepts like the Bures-Wasserstein distance. Its properties, such as the data processing inequality and the in-betweenness property, make it a robust and applicable divergence in various mathematical and physical contexts.

\subsection*{Metric Summary }
Table~1 summarizes the performance characteristics of various distance metrics applied to functional connectomes across parcellation scales. While geodesic-based methods such as the affine-invariant, log-Euclidean, and Bures-Wasserstein distances theoretically align with the non-Euclidean geometry of FCs, they often require careful tuning and are sensitive to the choice of regularization. In contrast, metrics like Alpha-z Bures Wasserstein offer a balance of strong performance and low tuning sensitivity, making them particularly suitable for robust FC comparison across varying conditions. Our proposed method builds upon these insights, addressing the limitations of conventional metrics by offering a more stable and accurate alternative that generalizes well across datasets and parcellation granularity.

\subsection*{QUANTIFICATION AND STATISTICAL ANALYISIS}

\subsubsection*{ Participant Identification}

Participant identification involves mapping an unknown participant’s data to one of the participants in the database. Since each task in the HCP data contains two runs for every participant, we used one run as training data (i.e., to form the database) and the other run for testing. Identification was performed on each condition (resting-state or task) separately. 

Participant identification is equivalent to $N$-class classification, where the objective is to label an individual’s FC matrix in the test data to one of the $N$ participants in the training data. To achieve this, we used a 1-Nearest Neighbor approach \cite{finn2015functional}. An FC matrix in the test data is labeled with the participant identity of the FC matrix that is most similar to it in the training data.

Suppose $Q_{\text{test}}^x$ is an unknown participant’s FC matrix. Then, the label of $x$ is given by:

\[
\text{label}(x) = \arg\min_{i=1}^{N} d(Q_{\text{train}}^i, Q_{\text{test}}^x)
\]

where $Q_{\text{train}}^i$ is the $i$th participant’s FC matrix in the training data, and $d(\cdot, \cdot)$ is a distance or similarity measure. Here, we compare the use of a different distance metric to a metric that gets from the Alpha-z-Bures Wasserstein divergence  measure.

\subsubsection*{Identification Rate Computation Algorithm}
\begin{algorithm}[H]
    \caption{Computation of Identification Rate between Functional Connectivity Matrices}
    \label{alg:IDRate}
    \textbf{Input:}
    \begin{itemize}
        \item $\mathcal{A} = \{A_i\}_{i=1}^{428}$: A set of functional connectivity (FC) matrices representing one session or condition.
        \item $\mathcal{B} = \{B_j\}_{j=1}^{428}$: A set of functional connectivity (FC) matrices representing another session or condition.
    \end{itemize}
    \textbf{Output:} Identification rate  between the two sets of FC matrices.
    
    \begin{algorithmic}
        \State \textbf{Compute Distance Matrix} $D_{AB} \in \mathbb{R}^{428 \times 428}$
        \For{$i = 1$ to $428$}
            \For{$j = 1$ to $428$}
                \State Compute distance $D_{AB}[i,j]$ between $A_i$ and $B_j$.
            \EndFor
        \EndFor
        
        \State \textbf{Prediction Using $\mathcal{A}$ as Training Set}
        \For{$j = 1$ to $428$}
            \State $\text{Pred}_A[j] \gets \arg\min_{i} D_{AB}[i,j]$
        \EndFor
        
        \State \textbf{Prediction Using $\mathcal{B}$ as Training Set}
        \For{$i = 1$ to $428$}
            \State $\text{Pred}_B[i] \gets \arg\min_{j} D_{AB}[i,j]$
        \EndFor
        
        \State \textbf{Compute Correct Identifications}
        \State $\text{CorrectIDs}_A \gets 0$, \quad $\text{CorrectIDs}_B \gets 0$
        \For{$i = 1$ to $428$}
            \If{$\text{Pred}_B[i] = i$}
                \State $\text{CorrectIDs}_B \gets \text{CorrectIDs}_B + 1$
            \EndIf
        \EndFor
        \For{$j = 1$ to $428$}
            \If{$\text{Pred}_A[j] = j$}
                \State $\text{CorrectIDs}_A \gets \text{CorrectIDs}_A + 1$
            \EndIf
        \EndFor
        
        \State \textbf{Compute Identification Rates}
        \State $\text{IDRate}_A \gets \frac{\text{CorrectIDs}_A}{428}$
        \State $\text{IDRate}_B \gets \frac{\text{CorrectIDs}_B}{428}$
        \State $\text{IDRate} \gets \frac{\text{IDRate}_A + \text{IDRate}_B}{2}$
        
        \State \Return $\text{IDRate}$
    \end{algorithmic}
\end{algorithm}

\subsubsection*{ Identification Accuracy}

Participant identification was performed using the first run as training data and the second run as testing data. For the $N$ participants in the testing data, accuracy was defined as:

\[
\text{Accuracy} = \frac{\text{Number of correctly labeled participants}}{\text{Total number of participants}}
\]

Then, the roles of the training and testing data were reversed, and accuracy was computed again. The reported identification accuracy was the mean of the two accuracy values.





\end{document}